\documentclass[aps,pra,superscriptaddress,twocolumn]{revtex4-2}
\usepackage{amsmath}
\usepackage{graphicx}	
\usepackage{bm} 
\usepackage{color}
\usepackage{setspace}
\usepackage{placeins}
\usepackage[dvipsnames]{xcolor}
\usepackage{mathtools}
\usepackage{hyperref}
\DeclarePairedDelimiter{\ceil}{\lceil}{\rceil}
\begin{document}
\title{First and second quantized digital quantum simulations of bosonic systems}
\date{\today}
\author{Mathias Mikkelsen}
\email[]{mathias@qunasys.com}
\affiliation{QunaSys Inc., Aqua Hakusan Building 9F, 1-13-7 Hakusan, Bunkyo, Tokyo 113-0001, Japan}
\author{Hubert Okadome Valencia}
\affiliation{QunaSys Inc., Aqua Hakusan Building 9F, 1-13-7 Hakusan, Bunkyo, Tokyo 113-0001, Japan}
 
\begin{abstract} 
We compare the basic resource requirements for first and second quantized bosonic mappings in a system consisting of $N$ particles in $M$ modes. In addition to the standard binary first quantized mapping, we investigate the unary first quantized mapping. Our comparison focuses on the $k$-body reduced density matrix ($k$-RDM) and two standard bosonic Hamiltonians. The first quantized mappings use less resources for off-diagonal terms of the $k$-RDM by a factor of $ \sim N^k$, compared to the second quantized mappings. The number of gates for the first quantized binary mapping increases faster with $M$ compared to the other mappings. Nevertheless, a detailed numeric analysis reveals that the binary first quantized mapping still requires fewer gates than the binary and unary second quantized ones for realistic combinations of $N$ and $M$, while requiring exponentially fewer qubits than the unary mappings.  Additionally, the number of CNOT and $R_z(\phi)$ gates necessary to express a single Trotter step of the Hamiltonian in the binary first quantized mapping is comparable to the (most efficient for a single Trotter step) unary first quantized one when $M = 2^n$ for both the Bose-Hubbard model and the harmonic trap with short-range interactions. Additionally the binary mapping leads to lower one-norms than the unary mapping making it the overall most efficient choice for qubitization-based quantum phase estimation.
\end{abstract}

\maketitle

\section{introduction}

Simulating complicated many-body quantum systems is a major motivation for the field of quantum computation \cite{Feynman1982}. A universal quantum computer is not required for quantum simulations, and analog simulators, in which Hamiltonians are specifically engineered using a high degree of control, have seen a lot of interest in the last 20 years using physical systems such as ultracold gases \cite{Bloch2008,Cazalilla2011,Jochim2012,Mistakidis2023}. However, expressing an arbitrary Hamiltonian in a language suited for a universal quantum computer is useful, as such devices continue to scale up. The majority of quantum computational research is based on binary systems, where the local basis has two degrees of freedom (so-called qubits). These systems have two main advantages: it is straightforward to develop a universal language of computation that can be described similarly to classical binary computation \cite{Nielsen_Chuang2010}, and there are many potential physical platforms for implementation \cite{Leon2021}. Although it is possible to implement quantum computations using a different local Hilbert space \cite{Wang2020,Ringbauer2022}, the development of qubit-based quantum computers is currently receiving the majority of efforts. 

Expressing many-body Hamiltonians using quantum logic gates defined on the qubit Hilbert space is therefore important. So far, a fermionic approach to the problem has been the focus of most attention, in particular to determine the electronic structure in quantum chemistry \cite{Cao2019,McArdle2020}. There are both theoretical and practical reasons for this: theoretically, the mapping of quantum chemistry problems to digital quantum computers is straightforward, as the second quantized fermionic Hamiltonian is described by a local two-level Hilbert space. This has led to steady progress in the theoretical development of a variety of quantum algorithms. Practically, quantum chemistry is of significant industrial interest \cite{McArdle2020}, and obtaining a quantum advantage in this field is a major milestone for quantum computation. The bosonic many-body problem has received less attention: although the first explicit construction of a second quantized mapping dates back to 2003 \cite{Rolando2003}, further studies have mostly been carried out within the past 5 years \cite{Sawaya2020a,trenev2023,majland2021,Kuwahara2024,peng2023,McArdle2019,Sawaya2019,Lotstedt2021,Sawaya2021,Sawaya2020b,bahrami2024,Liu2022,Tudorovskaya2024,Wang2023}. This includes comparative analyzes of different mapping methods \cite{Sawaya2020a,trenev2023}, optimizing resources through physically motivated approximations \cite{majland2021,Kuwahara2024}, and general error analysis for effective Hamiltonians \cite{peng2023}. As a lot of the interest in quantum computing comes from the point of view of quantum chemistry, problems related to the latter, such as vibrational spectra simulation \cite{trenev2023,McArdle2019,Sawaya2019,Lotstedt2021,Sawaya2021} have been the main focus, although more general simulations of, for example, Hubbard-like models have also been investigated \cite{Sawaya2020a,Kuwahara2024,peng2023}. 

While the majority of the work on mapping quantum many-body hamiltonians to digital quantum computers has focused on the second quantized mapping, the first quantized mapping was investigated as far back as the 1990s \cite{Abrams1997} and in recent years has seen an increased interest for the fermionic problem, particularly in the context of qubitization \cite{Yuan2021,Delgado2021,Rubin2023,Shokrian2023,berry2024,georges2024}. The main advantage of the first quantized mapping compared to the second quantized mapping is the ability to map an exponentially large number of modes. For a system with $M$ modes and $N$ particles, the first quantized mapping requires $N \cdot \ceil{\log_2{M}}$ qubits, while standard second quantized mappings such as Jordan-Wigner require $M$ qubits. A major disadvantage, however, is that although this mapping generally requires fewer qubits, a larger number of gates are necessary if we are mapping a system with the same number of modes/particles; the mapping of a single two-body off-diagonal element being upper bounded by $N^2 M^2$ compared to first order for the Jordan-Wigner mapping. For fermionic chemistry problems, being able to use larger basis sets, can make the first quantized mapping overall more gate-efficient \cite{Yuan2021} which explains the recent resurgence of interest. Conversely, for a small set of optimized atomic orbitals, the second quantized mapping can be more efficient and has consequently been the main focus for near-term quantum computation. A second disadvantage is the need for an explicit anti-symmetrization of the wavefunction. Fortunately, efficient anti-symmetrization algorithms exist \cite{Berry2018}, which turn its resource cost into a relatively insignificant part of a given algorithm.

Although, in principle, many of the  results for first quantized fermionic systems can be applied to bosonic systems, we focus on comparisons with second quantized bosons - an understudied area. Unlike fermions where the second quantized mapping is more resource-efficient for the same number of particles and modes, the second quantized bosonic mappings are naively less resource-efficient, although it depends on the particulars of the system and mapping. We therefore make a resource comparison between the unary first quantized (U1Q), binary first quantized (B1Q), unary second quantized (U2Q) and binary second quantized (B2Q) mappings. The U1Q mapping has not been studied before as it was deemed uninteresting for the fermionic case, but we show that it can have some advantages, compared to the second quantized mappings for the bosonic case. The first quantized mapping can lead to lower qubit and gate counts than the second quantized one, potentially allowing them to be implemented in noisy intermediate-scale quantum (NISQ) devices and early-fault tolerant quantum computers (FTQCs).

Symmetrizing a qubit representation of a given state is more complicated than anti-symmetrizing due to the existence of duplicate indices when multiple bosons occupy the same mode \cite{Berry2018}. However, there are a number of algorithms that can be implemented using a simple symmetrized initial state, namely $|N000...\rangle$ which corresponds to a single computational state. This computational state corresponds to the ground state of a non-interacting system when the basis functions for the Fock-space are the eigenstates of a single-particle (SP) or mean-field (MF) Hamiltonian. Any algorithm where the circuits obey bosonic symmetry applied to this state gives results in the bosonic symmetry sector. Some examples include time evolution which allows for quantum phase estimation (QPE) \cite{kitaev1995,Low2019,Kivlichan2020,Yuan2021}, or the recently proposed time-evolved quantum-selected configuration interaction (TE-QSCI) method \cite{mikkelsen2024,sugisaki2024,yu2025}.  Preparing better initial states for QPE, using adiabatic time evolution as investigated for Fermions in \cite{Guzik2005,Veis2014,Lee2023} can also be implemented using $|N000...\rangle$ as the initial state. Similarly variational quantum algorithms such as the variational quantum eigensolver (VQE) can be implemented using the Standard unitary coupled-cluster
with singles and doubles (UCCSD) ansatz \cite{Cederbaum2006,Tilly2021,Anand2022}, which also preserves the bosonic symmetry, although the circuit depth makes it difficult to implement for large system sizes on current NISQ hardware. An alternative is to use simpler symmetry-preserving variational circuits similar to a recent suggestion for the first quantized fermionic VQE \cite{horiba2023}. In principle a non-symmetry-preserving variational circuit should still find the bosonic ground state, as a standard particle-symmetric Hamiltonian of indistinguishable particles will obey bosonic symmetry, but it is likely less efficient, as it has more chance to get stuck in a local minima.

For states which do not have two bosons in the same mode, the method used for preparing the fermionic anti-symmetric states \cite{Berry2018} can be used. This makes it possible to prepare different seed states, which is useful for basis functions that are not SP or MF states, such as those needed in Bose-Hubbard models. Here, preparing the Mott-state  $|111...\rangle$, or charge-density-wave (CDW) state $|101...\rangle$ to use in conjunction with the algorithms suggested above makes more sense than using the simple $|N000...\rangle$ state. The time evolution of such initial states for different physical parameters is also an interesting simulation problem.

The paper is organized as follows. We briefly introduce the mappings, the reduced density matrix and the type of Hamiltonian of interest in section \ref{sec:Mappings}. We then introduce the unary and binary mappings of a generic Hilbert space in section \ref{subsec:genericHilbertspacemapping}. In sections \ref{subsec:2ndquantizedmapping} and \ref{subsec:1stquantizedmapping}, we detail the mappings of the second and first quantized operators discussing their mathematical form. In section \ref{sec:resource-comparison} we do a more detailed resource comparison.  Section \ref{subsec:RDM} focuses on the $k$-body reduced density matrix off-diagonal elements, while section \ref{subsec:Hamiltonians} investigates two paradigmatic bosonic model Hamiltonians  - the standard Bose-Hubbard model \cite{Cazalilla2011} (BHM) and the harmonic oscillator with short-range delta-interactions \cite{Jochim2012,Mistakidis2023} (HO). These sections investigate the number of CNOT and $R_z(\phi)$ rotation gates required to express the trotterized exponential of operators, the number of non-commuting Pauli groups in a given operator and the one-norm for the explicit Hamiltonians. Finally, we discuss the advantages and disadvantages of the various mappings and outline some future research directions in section \ref{sec:disussion}.

\section{Mappings, reduced density matrices and basic hamiltonians}
\label{sec:Mappings}
Second quantized mappings map the local $d$-level Hilbert space, which determines how many bosons occupy the site/mode, to qubits and bosonic creation/annihilation operators to Pauli operators. For a system with $M$ sites/modes, the mapping of the local Hilbert space is repeated for each site/mode. In contrast, the first quantized mappings assign a register to each of the $N$ bosons, storing the index of the site/mode it occupies, and therefore represent a fixed total-number sector. The latter is consequently less flexible, since it requires the total number of bosons to be conserved, but we will show that it is more efficient when mapping $k$-body reduced density matrices (
k$-RDMs$). That is, we can efficiently express the matrix elements of the $k$-RDM
\begin{align}
\hat{\rho}(l_1,...,l_{k},m_1,...,m_{k})=\prod_{j=1}^{k} \hat{a}_{l_j}^\dagger \prod_{j=1}^{k} \hat{a}_{m_j}.
\label{eq:reduceddensityoperator}
\end{align}
Here, $\hat{a}_{l_j}^\dagger, \hat{a}_{m_j}$ corresponds to the standard bosonic creation and annihilation operators, respectively. This also means that a typical Hamiltonian with two-body interactions, represented as
\begin{align}
\hat{H}=\sum_{kl} h_{kl} \hat{a}_k^\dagger \hat{a}_{l}+\frac{1}{2}\sum_{klmn} V_{klmn} \hat{a}_k^\dagger \hat{a}_{l}^\dagger\hat{a}_{m} \hat{a}_{n}
\label{eq:genericbosonicHamiltonian2nd}
\end{align}
in second quantization and as
\begin{align}
\hat{H}&=\sum_{\alpha=1}^N \sum_{kl} h_{kl} |\alpha,k\rangle \langle \alpha, l| \nonumber \\
&+\sum_{\alpha>\beta=1}^{N}\sum_{klmn} V_{klmn} |\alpha,k\rangle \beta,l \rangle \langle \beta,m | \langle \alpha, n |
\label{eq:genericbosonicHamiltonian1st}
\end{align}
in first quantization, with matrix elements determined by the single-particle Hamiltonian and interaction potential as 
\begin{align}
h_{kl} &= \int dx \phi_k^*(x) H_{sp} \phi_l(x), \\
V_{klmn}&= \int dx_1 dx_2 \phi_k^*(x_1) \phi_l^*(x_2) V(x_1,x_2) \phi_m(x_2) \phi_n(x_1)
\end{align}
can be expressed efficiently.

\subsection{Generic Hilbert space mapping}
\label{subsec:genericHilbertspacemapping}
Before presenting the first and second quantized mappings, it is useful to discuss the basic case of a generic Hilbert-space spanned by the basis $|0\rangle,...|d-1 \rangle$ and the operator 
\begin{align}
| l \rangle \langle m |.
\label{eq:basicoperator}
\end{align}
Note that each basis state is labeled using a unique number $l,m=0,1,...,d-1$. There are two elementary ways to map this problem. One is the unary mapping, which uses $d$ qubits. Here, the basis state number is determined by the index of the non-zero qubit, for example, if $d=6$,  $l=0$ and $l=2$ would be represented by $|0 \rangle = |100000 \rangle$ and $| 2 \rangle = |001000 \rangle$, respectively. The operator in Eq.(\ref{eq:basicoperator}) can be written in terms of the Pauli lowering and raising operators $\hat{S}_{l}^{\pm}=\frac{1}{2}(\hat{X}_l\pm i\hat{Y}_l)$ as 
\begin{align}
| l \rangle \langle m |=
  \begin{cases}
    \hat{S}_{l}^+ \hat{S}_{m}^-,        & \text{if } l \neq  m \\
   \frac{1}{2}(1-\hat{Z}_l)\delta_{lm}        & \text{if } l = m
  \end{cases}
\label{eq:unarymapping}
\end{align}
that is 4 Pauli strings each of length 2 for off-diagonal operators, a constant and a Pauli string for the diagonal operators.

The second way is a binary mapping.  Here, the index number is represented using $\ceil{\log_2(d)}$ qubits (i.e the smallest integer larger than $\log_2(d)$). Note that multiple binary mappings exist, for example the so-called Grey code was explored for second quantized bosons in \cite{Sawaya2020a}. However, the most commonly used binary mapping in quantum computation is the standard base 2 representation of the index number. In this paper we will focus only on this standard binary mapping. Here the qubit vectors are given by the binary representation of the number, i.e. $|l\rangle = |l_1...l_{\ceil{\log_2(d)}} \rangle$. The previous example would be represented by $\ceil{\log_2(6)}=3$ qubits, which for $l=0$ and $l=2$, is given by $|0 \rangle = |000 \rangle$ and $| 2 \rangle = |010 \rangle$, respectively.

Although it is more qubit-efficient, the binary mapping of the operator in Eq.(\ref{eq:basicoperator}) is more complicated and its implementation requires more quantum gates. It is given by 
\begin{align}
| l \rangle \langle m |= |l_1\rangle \langle m_1| \otimes ... \otimes |l_{\ceil{\log_2(d)}} \rangle \langle m_{\ceil{\log_2(d)}} | 
\end{align} 
with the relevant Pauli operators given by 
\begin{align}
|0 \rangle \langle 1 | &= \hat{S}^- , & |1 \rangle \langle 0 | &= \hat{S}^+ ,\nonumber\\
|0 \rangle \langle 0 | &= \frac{1}{2}(1+\hat{Z}) , & |1 \rangle \langle 1 | &= \frac{1}{2}(1-\hat{Z}).
\label{eq:compactpauli}
\end{align}
For any value $l\neq m$, this leads to $2^{\ceil{\log_2(d)}}$ Pauli strings. The number of gates required to express $|l \rangle \langle m |$ is correlated with the $l$, $m$ Hamming distance. For a Hamming distance of $d$ (when all bits are different), every Pauli string length is $\ceil{\log_2(d)}$. This corresponds to the maximum resource requirements of the operator in Eq.(\ref{eq:basicoperator}). For shorter Hamming distances, the Pauli strings will have varying lengths with a maximum of $\ceil{\log_2(d)}$.

\subsection{Second quantized mapping}
\label{subsec:2ndquantizedmapping}
The standard way of mapping bosons to qubits is based on representing the local $d$-level bosonic Hilbert space at each site/mode in terms of qubits, with $M$ registers corresponding to each mode. The second quantized site-local operators, such as $\hat{a}^\dagger_j,\hat{a}_j,\hat{n}_j$ are then mapped to Pauli operators.  If the local Hilbert space is described in terms of the states $|0\rangle,|1\rangle,...|d-1 \rangle$, where each state corresponds to a bosonic occupation of $0,1,..,d-1$, the creation, annihilation, and number operators can be formally defined in terms of the basis $|j,n\rangle$ where $j$ refers to the site/mode and $n$ to the local occupation as:
\begin{align}
\hat{a}^\dagger_j&= \sum_{n=0}^{d-2} \sqrt{n+1} |j,n+1\rangle \langle j,n |  \nonumber \\
\quad \hat{a}_j&= \sum_{n=0}^{d-2} \sqrt{n+1} |j,n\rangle \langle j , n+1|, \nonumber \\
 \hat{n}_j&= \sum_{n=1}^{d-1} n |j,n\rangle \langle j , n| .
\label{eq:basic2Qmapping}
\end{align}

For the unary mapping, the operators can be obtained by inserting the following relations in Eq. (\ref{eq:basic2Qmapping}):
\begin{align}
|j,n+1\rangle \langle j , n| &= \hat{S}_{j,n+1}^+ \hat{S}_{j,n}^- , \nonumber  \\
|j,n\rangle \langle j , n+1| &= \hat{S}_{j,n}^+ \hat{S}_{j,n+1}^-  \nonumber \\
|j,n\rangle \langle j , n|&=\frac{1}{2}(1-\hat{Z}_{j,n}).
\end{align}

In the binary mapping which expresses the states as $|n\rangle = |n_1...n_{\ceil{\log_2(d)}} \rangle$ we can map the creation, annihilation, and number operators by inserting 
\begin{align}
|j,n' \rangle \langle j , n| &= | j, n'_1...n'_{\ceil{\log_2(d)}} \rangle \langle j, n_1...n_{\ceil{\log_2(d)}}|  \nonumber \\
&=| j, n'_1 \rangle \langle j, n_1|\otimes ... \otimes|j,n'_{\ceil{\log_2(d)}}\rangle  \langle j, n_{\ceil{\log_2(d)}}|
\end{align}
in  Eq. (\ref{eq:basic2Qmapping}) with the relevant Pauli operators given by Eq.(\ref{eq:compactpauli}).

The creation, annihilation and density operators all require a sum over $d-1$ terms. On‑site interactions of the normal‑ordered form can be written as
\begin{align}
(\hat{a}_j^{\dagger})^k(\hat{a}_j)^k=\prod_{m=0}^{k-1}(\hat{n}_j-m) =\sum_{n=0}^{d-1} \prod_{m=0}^{k-1}(n-m) |j,n\rangle \langle j , n| .
\end{align}
Here it is clear that exactly $k$ values in this sum will always be zero i.e. whenever $n=m$, such that this sum always has $d-k$ terms.

For powers of the creation/annihilation operator, one can express the square (for the annihilation operator) as
\begin{align}
(\hat{a}_j)^2&= \sum_{n_1,n_2=0}^{d-2} \sqrt{n_1+1} \sqrt{n_2+1} |j,n_1\rangle \langle j , n_1+1 | \times \nonumber \\
&
j,n_2\rangle \langle j , n_2+1| \nonumber \\
&= \sum_{n_1,n_2=0}^{d-2} \sqrt{n_1+1} \sqrt{n_2+1} |j,n_1\rangle \langle j , n_2+1| \delta_{n_1+1,n_2} \nonumber \\
&=\sum_{n_2=1}^{d-2} \sqrt{n_2} \sqrt{n_2+1} |j,n_2-1\rangle \langle j , n_2+1|
\end{align}
and by induction, we see that one sum over $d-k$ terms is sufficient for the power $k$. The same can be shown for the creation operator. However, for sequences of the creation and annihilation operators on different sites, a sum over $(d-1)^k$ terms is required.  Similarly, a non-local density-density correlation operator like $\hat{n}_{m_1}...\hat{n}_{m_k}$ has $(d-1)^{k}$ terms in the sum. This is because the operators act on different local Hilbert spaces and no simplification beyond that which can be obtained for each local operator is possible.  Overall, the second quantized mapping is efficient at expressing site-local operators but inefficient at expressing non-local operators.

While the number of terms in these sums for the U2Q and B2Q mapping is the same, the terms are simpler for the U2Q mapping. Each term in the sum required for $\hat{a}^\dagger_j$ and $\hat{a}_j$ contains 4 Pauli strings of length $2$ in the U2Q mapping, while they contain $2^{\ceil{\log_2(d)}}$ Pauli strings of a maximum length $\ceil{\log_2(d)}$ for the B2Q mapping. When taking the sum in eq.(\ref{eq:basic2Qmapping}) for the unary mapping, all terms will contain unique Pauli strings. Conversely, for the binary mapping, the same Pauli strings can appear in multiple terms, reducing the effective number of Pauli strings required. How many times the same Pauli string appears strongly depends on the specific choice of $d$ and the number of Pauli strings is non-trivial. 

In the U2Q mapping, we can express the symmetric one-body density matrix with its hermitian conjugated (H.C.) as 
\begin{align}
&\hat{a}_l^\dagger\hat{a}_m + H.C. =\frac{1}{8}\sum_{n_1,n_2=0}^{d-2}\sqrt{n_1+1}\sqrt{n_2+1} \hat{S}_{lm}^{n_1,n_2} \nonumber\\
&= \frac{1}{8} \sum_{n_1,n_2=0}^{d-2}\sqrt{n_1+1}\sqrt{n_2+1} \bigg( \hat{X}_{l,n_1}\hat{X}_{l,n_1+1} \hat{X}_{m,n_2+1}\hat{X}_{m,n_2} \nonumber\\
&+ \hat{X}_{l,n_1}\hat{X}_{l,n_1+1} \hat{Y}_{m,n_2+1}\hat{Y}_{m,n_2} +\hat{Y}_{l,n_1}\hat{Y}_{l,n_1+1} \hat{X}_{m,n_2+1}\hat{X}_{m,n_2} \nonumber \\
&+\hat{Y}_{l,n_1}\hat{Y}_{l,n_1+1} \hat{Y}_{m,n_2+1}\hat{Y}_{m,n_2}+\hat{X}_{l,n_1}\hat{Y}_{l,n_1+1} \hat{Y}_{m,n_2+1}\hat{X}_{m,n_2} \nonumber \\
&+ \hat{Y}_{l,n_1}\hat{X}_{l,n_1+1} \hat{X}_{m,n_2+1}\hat{Y}_{m,n_2}-\hat{Y}_{l,n_1}\hat{X}_{l,n_1+1} \hat{Y}_{m,n_2+1}\hat{X}_{m,n_2} \nonumber \\
&-\hat{X}_{l,n_1}\hat{Y}_{l,n_1+1} \hat{X}_{m,n_2+1}\hat{Y}_{m,n_2}\bigg).
\end{align}
Here we obtain 8 Pauli strings of length 4 for each $\hat{S}_{lk}^{n_1,n_2}$ in the double sum. For comparison, 16 Pauli strings of length 4 are required for the non-symmetric version. This can be generalized to the $k$-RDM which requires $(4[d-1])^{2k}$ Pauli strings of length $4k$ for each off-diagonal term (ODT), while the symmetric version (i.e. the sum of the operator and its hermitian conjugate) requires half the number of Pauli strings. For the B2Q mapping, the number of Pauli strings has a complicated dependence on $d$ as previously explained, but the maximum number of Pauli strings required for the $k$-RDM ODTs is bounded by $2^{2k \ceil{\log_2(d)}}[d-1]^{2k}$ and the maximum length is bounded by $2k \ceil{\log_2(d)}$.

\subsection{First quantized mappings}
\label{subsec:1stquantizedmapping}
For the first quantized mapping, each boson is mapped to an individual site/mode with an index between $0,M-1$, with $N$ registers corresponding to each boson. Previous investigations \cite{Abrams1997,Yuan2021,Delgado2021,Rubin2023,Shokrian2023,berry2024,georges2024} have generally focused on binary mappings. For expressing the index in binary notation, $\ceil{\log_2(M)}$ qubits are required to encode the information, and the state for a given particle register $\alpha$ is expressed as
\begin{align}
| \alpha,l \rangle = |\alpha, l_1 l_2 ... l_{\ceil{\log_2{M}}} \rangle . 
\end{align}

Operators expressed in terms of $| \alpha, l \rangle \langle \alpha ,m |$ can be mapped to Pauli operators using Eq.(\ref{eq:compactpauli}). An alternative approach is the unary mapping. While this requires $M$ qubits rather than $\ceil{\log_2{M}}$ qubits for each boson, it generally requires fewer gates. 

Note that the U1Q mapping is physically equivalent to an SU(N) hardcore bosonic model restricted to the one-particle-per-flavor sector, with the unused states corresponding to different particle sectors. See the supplemental material of \cite{Mikkelsen2025} for a more detailed exposition of how SU(N) hardcore boson/fermion models in the one-particle-per-flavor sector relate to bosons. 

The symmetric $1$-RDM is given by 
\begin{align}
&(\hat{a}_l)^\dagger\hat{a}_m + H.C. = \sum_{\alpha=0}^{N-1} |\alpha , l\rangle \langle \alpha,m | +H.C.  \nonumber \\
{} &=  \sum_{\alpha=0}^{N-1} |\alpha,l_1 \rangle \langle \alpha,m_1) | \otimes \nonumber \\
&|\alpha, l_2 \rangle \langle \alpha, m_2| \otimes... \otimes |\alpha,l_{\ceil{\log_2(M)}} \rangle \langle\alpha,m_{\ceil{\log_2(M)}}|+H.C. ,
\end{align}
with the relevant Pauli operators given by Eq.(\ref{eq:compactpauli}) in the B1Q mapping, while the U1Q mapping allows for the explicit expression
\begin{align}
 \hat{a}_l^\dagger \hat{a}_m+H.C. &= \sum_{\alpha=0}^{N-1} | \alpha,l \rangle \langle\alpha, m |+H.C.  \nonumber \\
 &=  \sum_{\alpha=0}^{N-1} \hat{S}_{\alpha,l}^+ \hat{S}_{\alpha,m}^- +H.C.  \nonumber \\
 &= \sum_{\alpha=0}^{N-1}\left(\hat{X}_{\alpha,l}\hat{X}_{\alpha,m}+\hat{Y}_{\alpha,m}\hat{Y}_{\alpha,l}\right).
\end{align} 

In general, this requires a sum over $N$ terms. As in the second quantized case, the number and length of Pauli strings in each term is different for the binary and unary mappings. In the U1Q mapping, each term in the sum requires 4 Pauli (2 for symmetric) strings of length 2. For the B1Q mapping, however, $2^{\ceil{\log_2(M)}}$ Pauli strings with a maximum length of $\ceil{\log_2(M)}$ are required for each term. In general, the U1Q requires $2^{2 k} N^{k}$ strings of length $2k$ for the $k$-RDM (half for the symmetric case), while the number of strings required in the B1Q mapping is bounded by $N^{k}2^{k \cdot \ceil{\log_2(M)}}$, while the length of each is bounded by $k \cdot \ceil{\log_2(M)}$.

On‑site interactions of the normal‑ordered form can be expressed as
\begin{align}
 (\hat{a}_j^{\dagger})^k(\hat{a}_j)^k &= k!\sum_{\alpha_1>\alpha_2...\alpha_k=0}^{N-1} |\alpha_1,j \rangle \langle \alpha_1,j | \nonumber \\
 &| \alpha_2,j \rangle \langle \alpha_2, j | ... | \alpha_k,j \rangle \langle \alpha_k, j |
\end{align} 
which has $\frac{1}{k!}\prod_{j=0}^{k-1}(N-j)$ terms, where U1Q and B1Q are obtained by inserting Eq. (\ref{eq:unarymapping}) and Eq. (\ref{eq:compactpauli}), respectively. A non-local density-density correlation operator like $\hat{n}_{m_1}...\hat{n}_{m_k}$ needs $N^{k}$ terms.  Overall, the first quantized mapping is efficient in expressing ODTs, but less efficient for complicated local terms.

\begin{figure*}[htb!]
\centering
\includegraphics[width=1\linewidth]{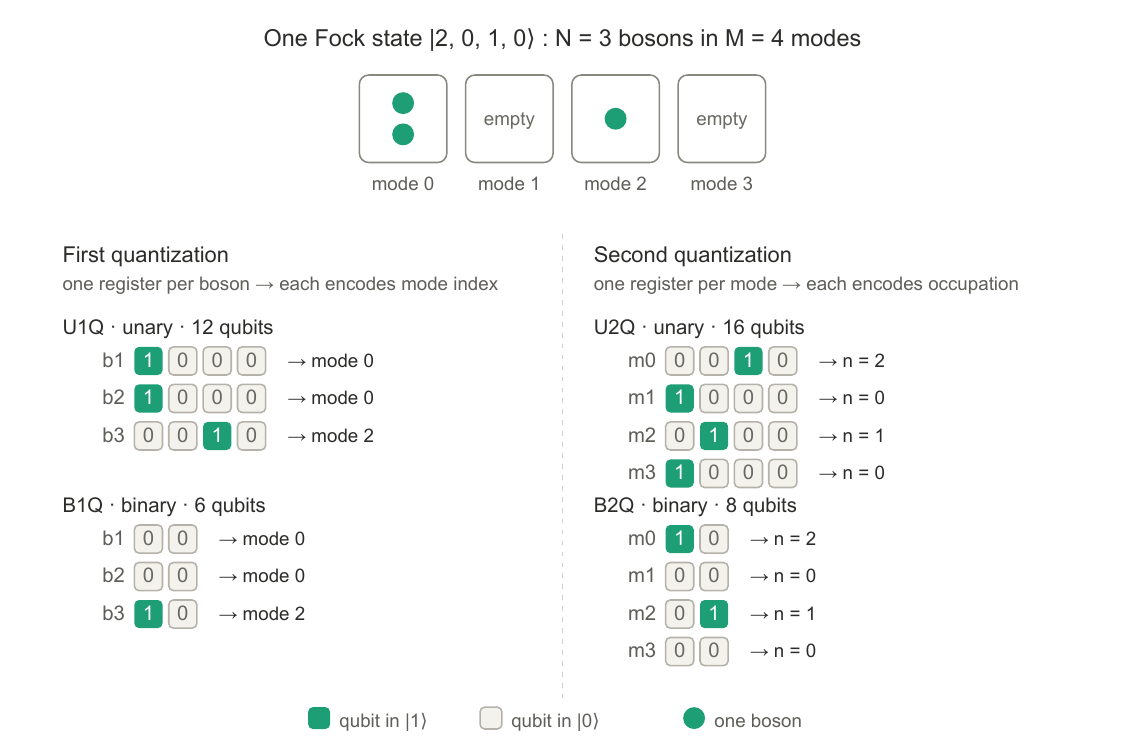}
\caption{An example of the binary and unary mappings for the first and second quantized version of the same Fock state.}
\label{fig:mappingvisualizations}  
\end{figure*}

In Table \ref{tab:mapingcomparison} the number of Pauli strings and their lengths for $k$-RDM ODTs for both the first and second quantized mappings are summarized. Fig.(\ref{fig:mappingvisualizations}) contains a visual illustration of how all mappings for a specific Fock state in a 3-particle 4-mode system is mapped to qubit registers.

Note that although we used a sequential register corresponding to each boson, it is straightforward to consider a different ordering for the unary mapping. For example, one could use the first $N$ qubits to express the information about whether a particle exists at index 0 for each boson, the next N to express whether a particle exists for index 1 etc. For all-to-all connectivity, which is the focus of the resource comparisons in this paper, this ordering is irrelevant, but given a specific connectivity and model, optimizing the ordering can become important.  

\begin{table*}[t]
\begin{tabular}{|l|l|l|l|l|}
\hline
                     & U1Q &  U2Q & B1Q & B2Q \\
 \hline
qubits & $N \cdot M$ &  $M \cdot (N+1)$ & $N \cdot \ceil{\log_2(M)}$ & $M \cdot \ceil{\log_2(N+1)}$    \\
 \hline
 Pauli strings &   $2^{2 k} N^{k}$  &  $(4N)^{2k}$   &  $2^{k \cdot \ceil{\log_2(M)}}N^{k}$ & $2^{2k \ceil{\log_2(N+1)}}N^{2k}$ \\
 \hline
 Pauli string length & $2k$ & $4k$ & $k \cdot \ceil{\log_2(M)}$ & $2k \ceil{\log_2(N+1)}$ \\
 \hline
\end{tabular}
\newline
\newline
\caption{Summary of the number of qubits required for each mapping and the Pauli strings/lengths required for the $k$-RDM ODT elements. For the unary mappings the formulas correspond to the exact number of Pauli strings and their exact lengths, but for the binary mappings they correspond to upper bounds and the actual values can be much lower.}
\label{tab:mapingcomparison}
\end{table*}

\section{Resource comparisons}
\label{sec:resource-comparison}
In addition to considering the Pauli strings and their length required for mapping the $k$-RDM, we consider a more concrete resource estimation of the number of gates required to implement a single Trotter step of their complex exponential. This involves the evaluation of exponentials of Pauli strings $e^{- i t w_l \prod_j \sigma_{j,k_j}}$, where $j$ indicates the lattice/mode index and $k_j$ the type of Pauli operator $(X,Y,Z)$. We consider the following elementary gates: the single qubit $z$-rotation gate $R_z(\phi)$ with arbitrary angle $\phi$, the CNOT gate, and single-qubit Clifford gates. The number of CNOT gates is more important than the number of single-qubit gates for NISQ devices, as the fidelity of the former is worse than the latter. However, the number of $R_z(\phi)$ gates becomes important, while the number of CNOT gates becomes marginal for (early-) FTQC devices where Clifford gates, including CNOT, are relatively cheap to implement, while non-Clifford gates such as $R_z(\phi)$ are the most cost-consuming. Representing each exponential via a basic CNOT staircase, the exponential of a Pauli string of length $p$ can be expressed using $n_{\text{CNOT}} = 2(p-1)$ CNOT gates and 1 $R_z(\phi)$ gate \cite{Sawaya2020a}. These comparisons are useful for QPE and time-evolution, but are also important for determining the resource requirements of the UCCSD-ansatz used in VQE. Finally we investigate the number of bitwise commuting Pauli (BWCP) groups, which is important for VQE calculations to accurately estimate the expectation value of an operator. To compare the first and second quantized mappings for $N$ particles in $M$ modes, a local Hilbert space of $d=N+1$ is chosen for the second quantized mappings, to take into account the full $N$-particle subsector of the Hilbert-space. 

We investigate the $k$-RDM to obtain a schematic understanding of how the mappings perform in terms of simple gate counts in section \ref{subsec:RDM}, but Hamiltonian operators consist of sums over such elements. Depending on the specifics of the one- and two-body interaction tensors, the various mappings can potentially lead to further simplification. In particular, the B1Q mapping, where multiple one- and two-body terms are mapped to the same single and pairs of particle registers, is likely to allow for simplification. A full resource analysis for e.g. QPE requires an actual Hamiltonian for meaningful comparisons.

In section \ref{subsec:Hamiltonians} we consider two paradigmatic bosonic models, the BHM and the HO with short-range interactions.
The BHM has a Hamiltonian given by 
\begin{align}
\hat{H}_{\text{BH}}&= -J \sum_{j} \hat{a}_j^\dagger \hat{a}_{j+1}+H.C. + U/2 \sum_{j} \hat{a}_j^\dagger \hat{a}_{j}^\dagger\hat{a}_{j} \hat{a}_j 
\label{eq:BHM}
\end{align}

while the HO with short-range interactions is given by Eqs.\ref{eq:genericbosonicHamiltonian2nd},\ref{eq:genericbosonicHamiltonian1st}, with matrix elements determined by 
\begin{align}
H_{kl} &= \hbar \omega (l+\frac{1}{2}) \delta_{k,l}, \\
V_{klmn}&= g \int dx \phi_k^*(x) \phi_l^*(x) \phi_m(x) \phi_n(x),
\end{align}
where $\phi_n(x)$ are the eigenstates of the non-interacting harmonic oscillator.

First we consider the resource requirements for Trotterization which is an extension of the resource analysis done for the $k$-RDM. However, in recent years an alternative implementation of QPE utilizing block-encoding has also been the subject of much attention \cite{Berry2019,Yuan2021,Delgado2021,Rubin2023,Shokrian2023,berry2024,georges2024}. In this encoding the expensive non-Clifford Toffoli gates required scales with the square root of the number of coefficients in the linear sum of unitaries (LCU) representing a Hamiltonian and its one-norm, which we therefore investigate as well. 

\subsection{Gate counts for $k$-RDM and density interactions}
\label{subsec:RDM}
For the unary mappings, the number of gates can be estimated precisely based on the analytic formulas for the number of Pauli strings and their length presented in section \ref{sec:Mappings}. The number of $R_z(\phi)$ rotation gates required to express the exponential is directly equivalent to the required number of Pauli strings. The $k$-RDM ODTs requires $2^{2 k} N^{k}$ Pauli strings of length $2k$ in U1Q mapping and $(4N)^{2k}$ Pauli strings of length $4k$ in the U2Q mapping. This means that $k$-RDM ODTs are expressed with a factor of $(4 N)^k$ fewer Pauli strings (their exponential forms require fewer rotation gates by the same factor) in the U1Q mapping. Expressing the exponential strings for a Trotter step then requires $2(2k-1)2^{2k} N^k$ CNOT gates for the U1Q mapping and $2(4k-1) (4N)^{2k}$ CNOT gates for the U2Q mapping. The U1Q mapping can therefore express the exponential string using a factor of
\begin{align}
\frac{n_{\text{CNOT,U2Q}}}{n_{\text{CNOT,U1Q}}}=\frac{4-\frac{1}{k}}{2-\frac{1}{k}}(4N)^k
\label{eq:kbodyratio}
\end{align}
fewer CNOT gates than the U2Q mapping.

This is done under the assumption that we want to investigate the exact same model. However, the U2Q mapping has an advantage compared to the U1Q mapping: one can choose the local Hilbert space such that $d<N+1$ and still get a physically meaningful result. However, this introduces errors due to truncation. For $d<N+1$, the number of Pauli strings or CNOT gates can be estimated by replacing $N$ with $d-1$ for the U2Q.  To express the off-diagonal elements of the $k$-RDM with the same number of CNOT gates as the U2Q mapping one needs to solve $\frac{n_{\text{CNOT,U1Q}}}{n_{\text{CNOT,U2Q}}}=1$ for $d$ in terms of the number of particles $N$, which leads to
\begin{align}
d = \frac{\sqrt{N}}{2} \left( \frac{2k-1}{4k-1} \right)^{\frac{1}{2k}} + 1
\end{align}
The equivalent formula for the number of Pauli strings is a bit simpler and independent of $k$ giving 
\begin{align}
d = \frac{\sqrt{N}}{2} + 1
\end{align}
which is also the limit of the CNOT expression as $k \rightarrow \infty$.
For a large $N=100$ system, this gives $3.9, 5.0, 5.4,6$ for $k=1,2,3$ and the $k$-independent d, respectively, which means a ($d-1$) truncation of $N_\text{max}=3,4,5$. For a moderate $N=10$ number of particles, $d=1.9, 2.3, 2.4,2.6$ which means a truncation of $N_\text{max}=1$ (hardcore bosons) or $N_\text{max}=2$ if we allow for using slightly more resources. For the simple BHM and $U=2J$ exact diagonalization for $N=10,M=11$ results in a ground state energy of $-14.28J$. Hardcore bosons physically correspond to $U/J=\infty$ and have the same spectrum as free fermions with the ground state energy given by $-2J$. Although this is strictly required for equivalent resource use for $N=10$, it is clear that this truncation is not physical, so we consider diagonalization in the restricted basis $N_\text{max}=2$ which results in $-13.01J$, which is still quite wrong. In practice, it is therefore not possible to get accurate results with the same number of resources, in particular for models that do not have strong on-site repulsion terms. 

Due to the particle exchange symmetry in the first quantized mapping, it is sufficient to measure the expectation value for an arbitrary particle register $ \langle |\alpha , l\rangle \langle \alpha,m | \rangle$ for the $1$-RDM, while the $2$-RDM requires measurement of $\langle |\alpha,k\rangle | \beta,l \rangle \langle \beta,m | \langle \alpha, n | \rangle$ for two arbitrary registers (one arbitrary register $\alpha=\beta$ and one arbitrary register $\alpha \neq \beta$). Similar considerations are valid for higher-order $k$-RDMs.  However, this is less robust against errors and a measurement can be carried out on all registers in parallel using a similar number of measurements because Pauli operators in different particle registers always commute. The second quantized mappings similarly allow the measurement of creation/annihilation operators at different sites to be done in parallel, resulting in a small number of successive measurements. The $k$-RDM ODTs results in $2^{2k}$ BWCP groups for the U1Q mapping and $4^{2k}$ BWCP groups for the U2Q mapping, such that the number of groups in the U1Q is a factor of $(2)^{2k}$ less.   

In terms of resource efficiency, the U1Q mapping is particularly good at expressing off-diagonal $k$-RDM terms. However, models can contain density terms and density-density correlations.  Estimating the resource use for these types of terms is therefore also important.

For density-density interactions between different $\hat{n}_{j_1}....\hat{n}_{j_k}$ sites, the U2Q mapping requires $\sum_{p=1}^k\binom{k}{p}N^p$ Pauli strings where $p$ is the length of the string for each different term in the sum ($p$ takes values $p=1,2,..,k$). This is because each $\hat{n}_{j}$ is a sum of $N$ single-$\hat{Z}$ terms (plus a constant identity piece), so in the product over the $k$ sites a Pauli string of length $p$ arises whenever $p$ of the sites contribute one of their $\hat{Z}$ terms and the remaining $k-p$ contribute the identity, giving $\binom{k}{p}$ ways to choose which sites carry a $\hat{Z}$ and $N$ index choices on each, hence $\binom{k}{p}N^{p}$ strings of length $p$. This is identical to the U1Q mapping, which requires $\sum_{p=1}^k\binom{k}{p}N^p$ non-identity Pauli strings of varying length $p$, using the same principle. As the Pauli strings only contain $\hat{Z}$, they can all be measured simultaneously in both cases. Consequently, for these terms, the two mappings are equally efficient. 

For on-site normal ordered interactions $ (\hat{a}_j^{\dagger})^k(\hat{a}_j)^k$, the U2Q mapping only needs $N-k+1$ Pauli strings of length 1. This means that no CNOT gates are required to express the Trotter step operator. However, the U1Q mapping requires $\frac{1}{m!}\prod_{j=0}^{m-1}(N-j)$ Pauli strings of length $m$, where $m=1,..,k$. This follows because each $|\alpha_i,j\rangle\langle\alpha_i,j|=\frac{1}{2}(1-\hat{Z}_{\alpha_i,j})$ acts on a distinct particle register, so on expanding the product and combining terms a Pauli string of length $m$ corresponds to choosing which $m$ of the $N$ registers carry a $\hat{Z}$, giving $\binom{N}{m}=\frac{1}{m!}\prod_{j=0}^{m-1}(N-j)$ strings of length $m$. So for e.g. $k=2$, $\frac{N(N-1)}{2}$ strings of length 2 and $N$ strings of length 1 is required, such that the number of $R_z(\phi)$ rotation gates is $\frac{N(N+1)}{2}$, while the number of CNOT gates is $N(N-1)$.  As the Pauli strings only contain $\hat{Z}$, they can be measured simultaneously in both mappings. Such on-site powers of the density are the only terms for which the U1Q mapping is less efficient. 

\begin{figure*}[htb!]
\centering
\includegraphics[width=1\linewidth]{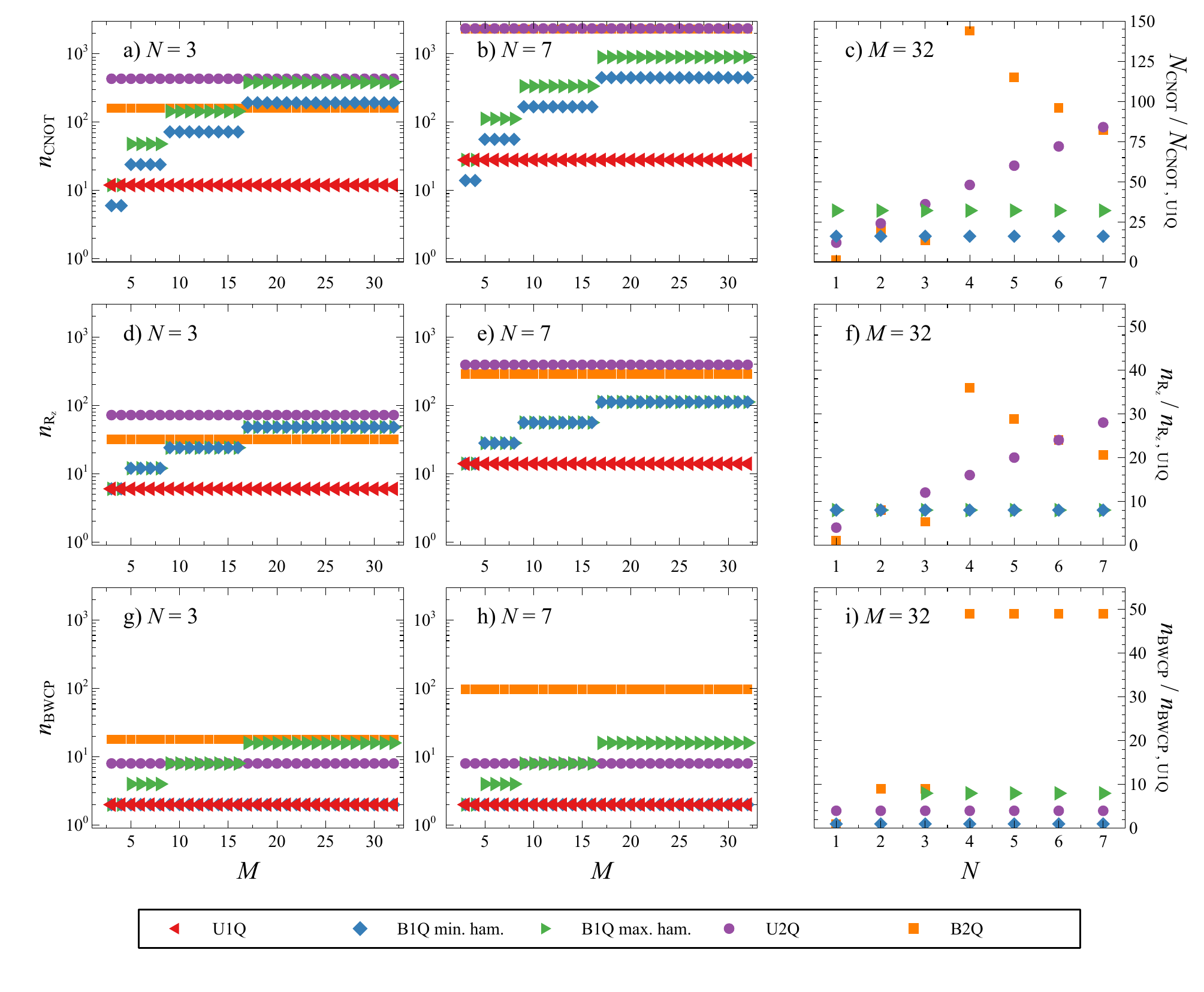}
\caption{Resource comparisons for symmetric  $1$-RDM ODTs $\hat{a}^\dagger_k\hat{a}_{l}+\hat{a}^\dagger_{l}\hat{a}_k$. Aside from the B1Q mapping all calculations are for $\hat{a}^\dagger_0\hat{a}_1+\hat{a}^\dagger_1\hat{a}_0$. For the B1Q mapping a calculation for this element corresponding to the minimum Hamming distance and one at indices corresponding to the maximum Hamming distance are both shown. (a)-(c) corresponds to the number of CNOT gates required to express one Trotter step of it's exponential, with (a) and (b) as a function of $M$ for $N=3$ and $N=7$ respectively, while (c) shows the number of CNOT gates normalized to the number required for the U1Q mapping as a function of $N$ for $M=32$. (d)-(f) shows the number of $R_z(\phi)$ gates required for the exponential for the same physical parameters.  (g)-(i) shows the number BWCP groups for the same parameters.}
\label{fig:1RDMresourcecomparison}  
\end{figure*} 

\begin{figure*}[htb!]
\centering
\includegraphics[width=1\linewidth]{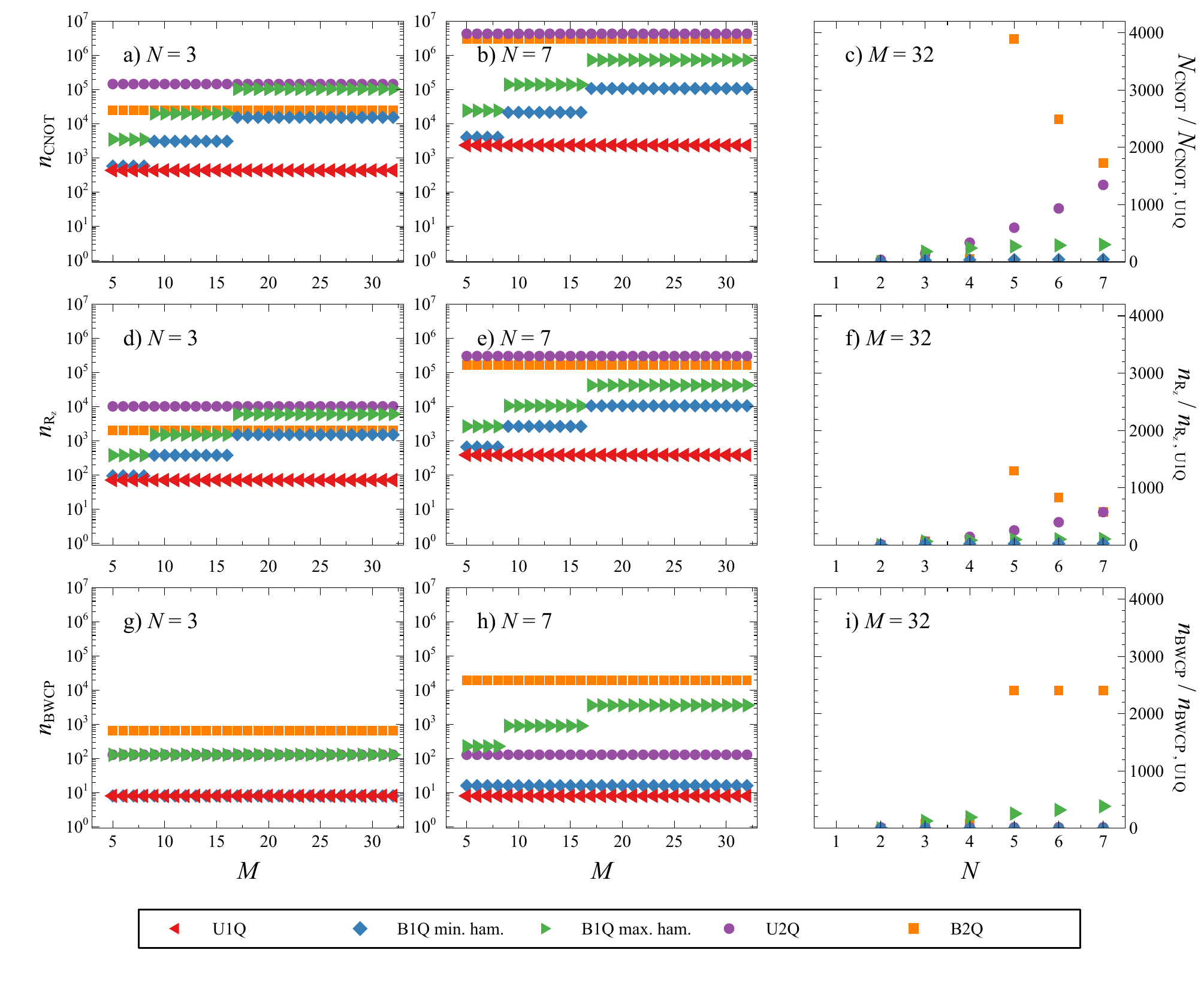}
\caption{The same plots as in Fig.\ref{fig:1RDMresourcecomparison}, but for symmetric $2$-RDM ODTs  $\hat{a}^\dagger_k \hat{a}^\dagger_l \hat{a}_m \hat{a}_n+\hat{a}^\dagger_n \hat{a}^\dagger_m \hat{a}_l \hat{a}_k$. Aside from the B1Q mapping all calculations are for $\hat{a}^\dagger_0 \hat{a}^\dagger_2 \hat{a}_1 \hat{a}_3+\hat{a}^\dagger_3 \hat{a}^\dagger_1 \hat{a}_2 \hat{a}_0$. For the B1Q mapping a calculation for this element corresponding to the minimum Hamming distance and one at indices corresponding to the maximum Hamming distance are both shown. }
\label{fig:2RDMresourcecomparison}  
\end{figure*} 

In the binary mappings, multiple terms will result in the same Pauli string and can be collected in a single term. Due to the complex nature of the mapping, it is not straightforward to account for this and a simple analytic comparison is not possible. We therefore investigate the number of required CNOT, $R_z(\phi)$ gates, and BWCP groups by implementing the mappings numerically, counting the number of CNOT and $R_z(\phi)$ gates required to express the resulting exponential strings and using the Quri-Parts software \cite{quri-parts} to find the number of BWCP groups.  In Figs.(\ref{fig:1RDMresourcecomparison},\ref{fig:2RDMresourcecomparison}) the number of CNOT and $R_z(\phi)$ gates required for one Trotter step of the exponential and the number of BWCP groups are shown for the symmetric $1$-RDM and $2$-RDM ODTs, respectively. 

For the B1Q mapping, the specific choice of indices for ODTs is important, as it impacts the number of gates. We consider indices corresponding to the minimum and maximum possible Hamming distances.

The U1Q uses the least amount of resources in general. This includes both the number of CNOT gates, $R_z(\phi)$ gates, and BWCP groups. As expected, the second quantized mappings are increasingly inefficient for large $N$ because the number of required gates grows polynomially faster with $N$. Although the B1Q mapping requires substantially more CNOT and $R_z(\phi)$ gates than the U1Q mapping their ratio is a constant as a function of $N$. For $N=3$ the U2Q mapping can be more efficient than the B1Q mapping depending on $M$ and the Hamming distance, but for $N=7$ it always uses more CNOT and $R_z(\phi)$ gates, as displayed for $M\leq 32$ in the figures.

The binary mappings lead to a much larger number of BWCP groups (except for the minimum Hamming distance B1Q case) and require more measurements to determine the expectation values. It is less clear that the number of BWCP groups scales favorably with $N$ for the B1Q mapping, being very dependent on the specific indices of the matrix element, although even for indices with the maximum Hamming distance, the scaling with $N$ improves compared to the B2Q mapping. Conversely, the scaling with $M$ is worse for the B1Q mapping compared to the B2Q mapping. 

\subsection{Algorithmic performance for physical models: Bose-Hubbard model and harmonic oscillator}
\label{subsec:Hamiltonians}

In this section we consider the BHM Hamiltonian and the HO Hamiltonian with short-range interactions. 
\subsubsection{Trotterization}
We investigated the Trotter error for small versions of these models in the supplemental material. Although naive worst error estimations make the first quantized mappings look much more accurate for a fixed number of steps due to the much smaller number of Pauli strings the actual trotter error is pretty similar across the different mappings, perhaps slightly better for the first quantized ones. The important measure is therefore the resource use per trotter step which we explore in detail in this section.

In the BHM the hopping terms are more efficiently expressed in the U1Q mapping, but the interaction terms are more efficiently expressed in the U2Q mapping. For periodic boundary conditions (PBC) the BHM has $M$ symmetric hopping terms when $M>2$ (for $M=2$ the extra terms due to PBC will be copies of the existing terms). This  means that the U1Q mapping for PBC uses 
\begin{align}
n_{\text{RZ,U1Q}}&= 2 M N +\frac{MN(N+1)}{2} \\
n_{\text{CNOT,U1Q}}&= 4MN +M N(N-1)
\end{align} 
while the U2Q mapping uses 
\begin{align}
n_{\text{RZ,U2Q}}=& 8 M N^2+M(N-1) \\
n_{\text{CNOT,U2Q}}&= 48MN^2.
\end{align}
This leads to a ratio between the two of
\begin{align}
\frac{n_{\text{RZ,U2Q}}}{n_{\text{RZ,U1Q}}} &= 16-\frac{78N+2}{N(N+5)} \\
\frac{n_{\text{CNOT,U2Q}}}{n_{\text{CNOT,U1Q}}} &=48-\frac{144N}{N(N+3)}
\end{align}
which has values of $R_z(\phi)$ (CNOT) ranging from $4.7$ ($19.2$) at $N=2$ to an asymptotic value of $16$ ($48$) depending on $N$ (for a moderate number of particles $N=10$, the ratio is respectively $10.8$ and $36.9$). 

\begin{figure*}[htb!]
\centering
\includegraphics[width=1\linewidth]{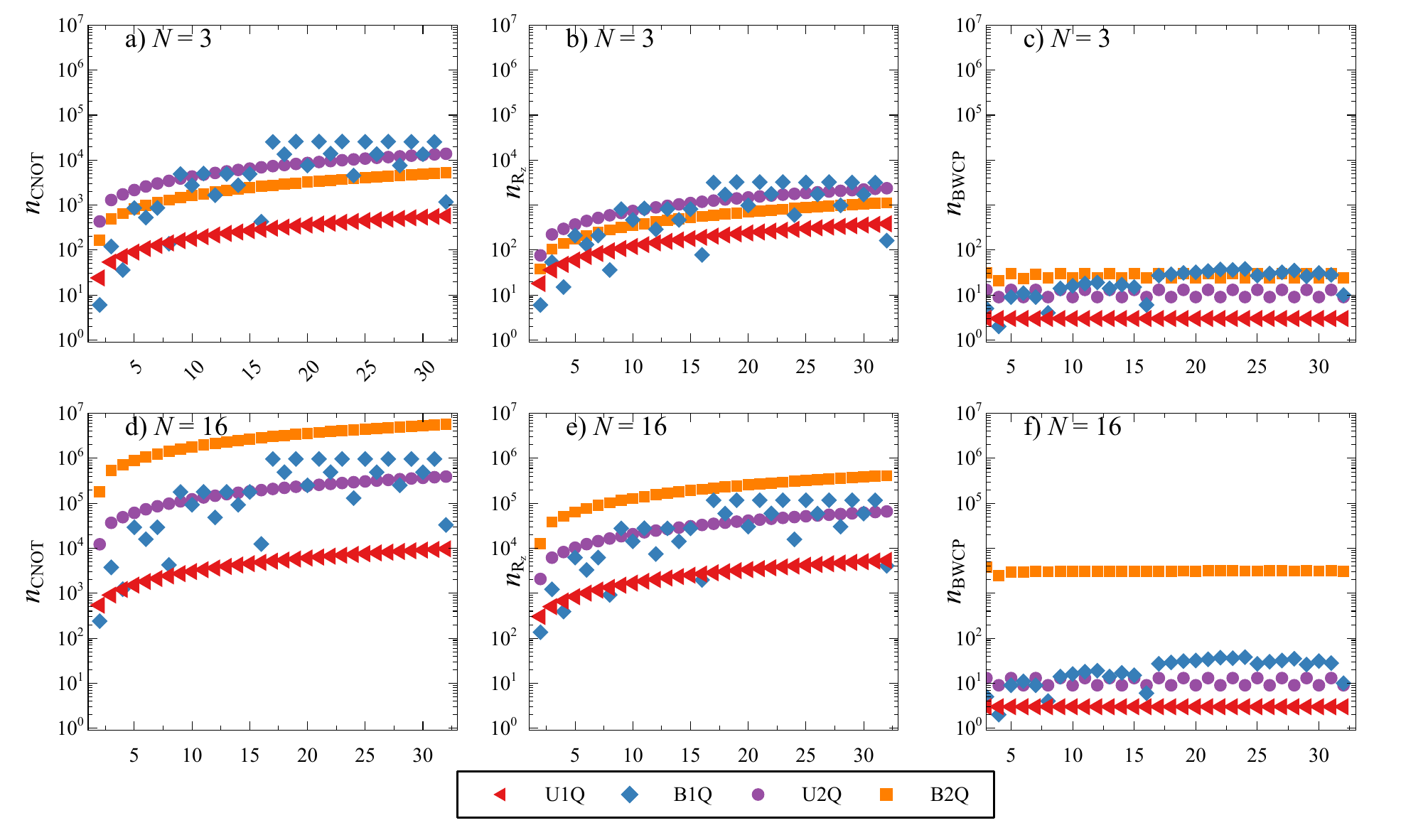}
\caption{Resource comparisons for the BHM as a function of $M$. (a),(b),(c) corresponds to $N=3$ while (d),(e),(f) corresponds to $N=16$.   (a,d) corresponds to the number of CNOT gates required to express one Trotter step of the exponential of the Hamiltonian, while (b,e) shows the corresponding number of $R_z(\phi)$ rotation gates, and (c,f) shows the number of BWCP groups.}
\label{fig:BHMresourcecomparison}  
\end{figure*} 

\begin{figure*}[htb!]
\centering
\includegraphics[width=1\linewidth]{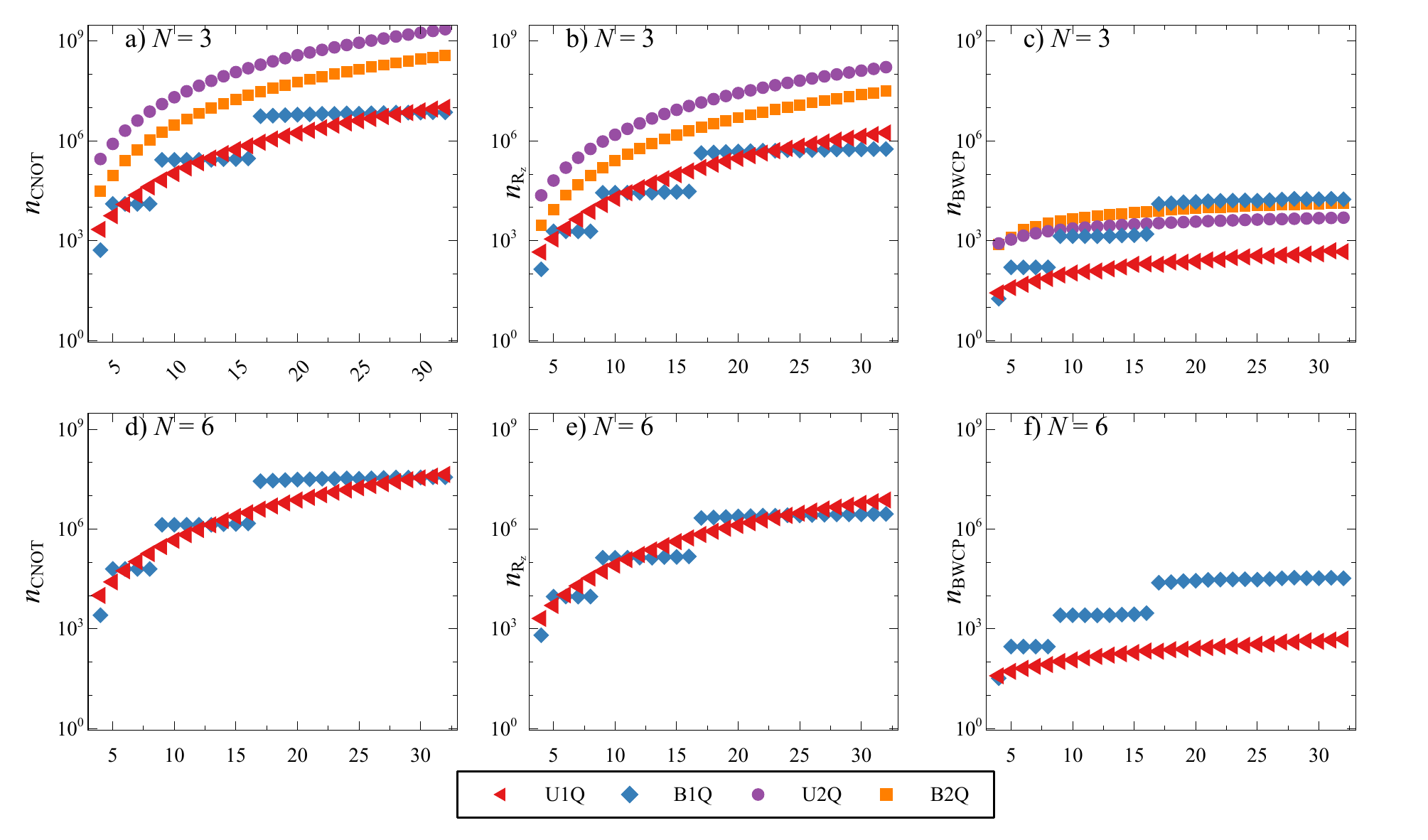}
\caption{Resource comparisons for the HO as a function of $M$. (a),(b),(c) corresponds to $N=3$ while (d),(e),(f) corresponds to $N=6$.   (a,d) corresponds to the number of CNOT gates required to express one Trotter step of the exponential of the Hamiltonian, while (b,e) shows the corresponding number of $R_z(\phi)$ rotation gates which also corresponds to the total number of Pauli strings in the LCU, and (c,f) shows the number of BWCP groups.}
\label{fig:HOresourcecomparison}  
\end{figure*}

\begin{figure*}[htb!]
\centering
\includegraphics[width=1\linewidth]{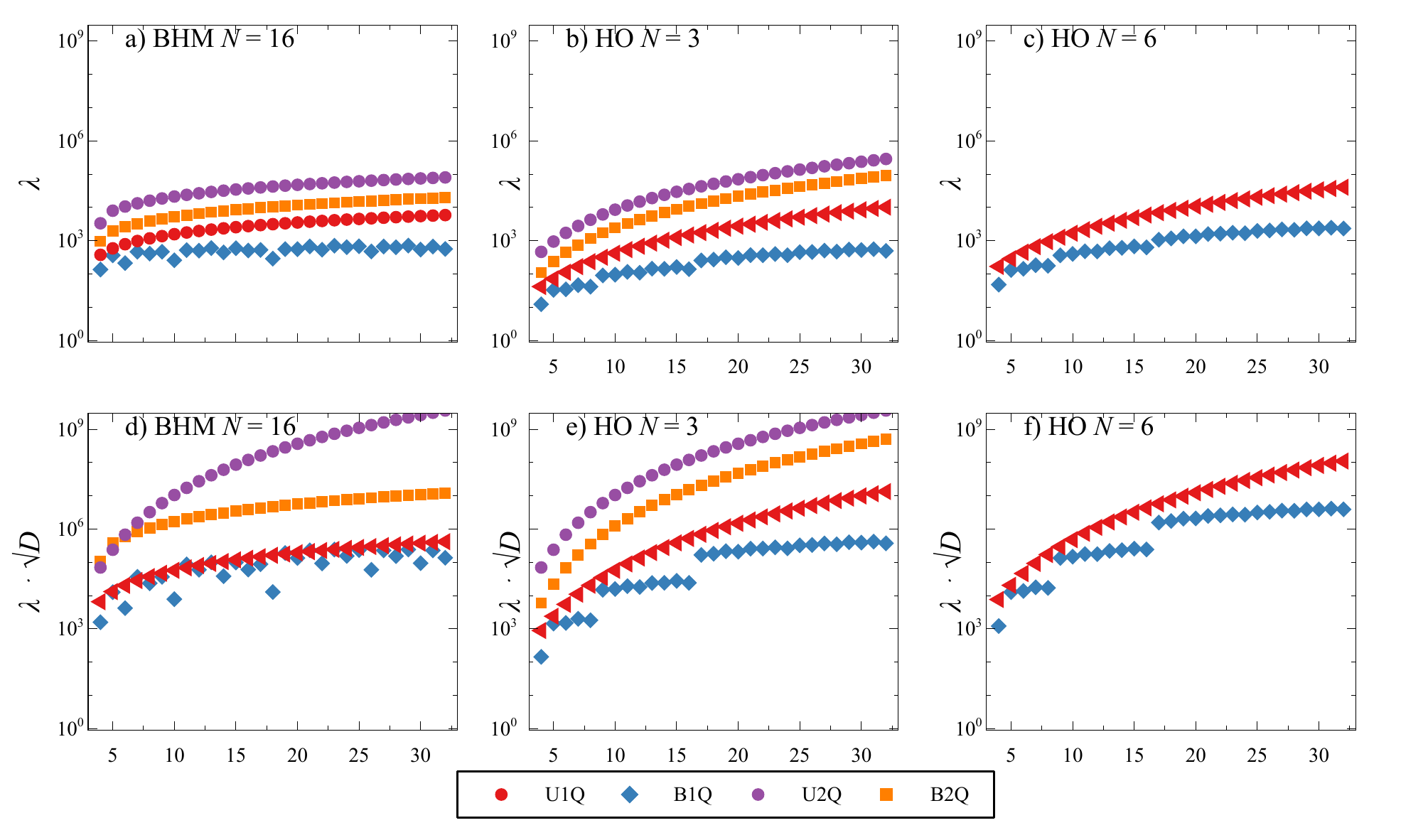}
\caption{Resource comparisons for the BHM and HO as a function of $M$. (a),(b),(c) corresponds to the one-norm $\lambda$, while (d),(e),(f) corresponds to $\lambda \cdot \sqrt{D}$, where D is the number of Pauli strings in the LCU.}
\label{fig:Onenormcomparison}  
\end{figure*}

From Fig. \ref{fig:BHMresourcecomparison} (a, b, d,e), the numerical evaluation agrees with the analytic estimate, leading to a constant factor improvement for U1Q compared to the U2Q mapping; with $R_z(\phi)$ (CNOT) ratios ranging from $6.2$ ($24$) to $12.3$ ($40.4$) for $N=3$ and $N=16$, respectively. The B2Q mapping uses comparatively more resources for $N=16$ than for $N=3$. Interestingly, the B1Q mapping is particularly efficient when $M=2^n$; using a number of CNOT and $R_z(\phi)$ gates that are closer to the U1Q mapping. It also leads to fewer BWCP groups than $M \neq 2^n$ as seen in (c,f).  This is related to terms canceling out when all bit values of a certain length are represented in the lattice, as also observed for the B2Q mapping in \cite{Sawaya2020a}. For our results to be generalizable to other models, we used a naive implementation of the Trotter operator, even though a more efficient representation is possible in the B1Q mapping for the nearest-neighbor Hubbard Hamiltonian, as explained in \cite{Abrams1997}. Note that the number of CNOT gates required for $M=32$, $N=3$ ($N=16$) is on the order of $10^3$ ($10^4$) in the first quantized mappings and may be implementable on near-term NISQ hardware, while the second quantized mappings are clearly not. Pushing to larger sizes in NISQ systems seems quite challenging, however. In contrast, the number of $R_z(\phi)$ gates required is also on the order of $10^4$  for $M=32$, $N=16$  which is well within the current maximum gate count in the recently proposed early-FTQC STAR architecture \cite{Akahoshi2024} where larger systems sizes are therefore more within reach.

In Fig. \ref{fig:HOresourcecomparison} the number of CNOT and $R_z(\phi)$ gates required to express the exponential of the HO Hamiltonian as well as  the number of BWCP groups required to measure its expectation value are shown as a function of $M$ for $N=3,6$. Due to the large number of gates involved in the second quantized mapping as illustrated by the $2$-RDM results in Fig.\ \ref{fig:2RDMresourcecomparison}, obtaining results for large values of $M$ becomes numerically difficult as $N$ increases. We therefore only compare the first quantized mappings for $N=6$, while we compare all for $N=3$. Already for $N=3$, the U2Q mapping uses between 106 and 190 (130 and 223) times more $R_z(\phi)$ (CNOT) gates than the U1Q mapping, while the B2Q mapping uses between 12 and 31 (14 and 36) times more, depending on the $M$ values. The reason why the B2Q mapping is more efficient than the U2Q mapping is due to the small value of $N=3$ consistent with previous results. For larger $N$, it becomes less efficient than the U2Q mapping. Once again, we find that the B1Q mapping is most efficient for $M=2^n$, actually outperforming the U2Q mapping. However, contrary to the BHM, this is due to the growth in the number of gates required for the unary mapping, while it is almost constant relative to the number of bits in the binary mapping. The number of BWCP groups is always the smallest for the U1Q mapping, with the relative ranking of the remaining mappings dependent on the specific values of $M$ and $N$. The exact same tendencies are observed in the first quantized mapping for both $N=3$ and $N=6$. For the HO which contains many $2$-RDM ODTs implementing the time evolution circuit on NISQ hardware is unrealistic due to the large number of CNOT gates, while the number of $R_z(\phi)$ gates required are still beyond near-future hardware proposals such as STAR \cite{Akahoshi2024}. Simulating bosonic problems of this type therefore relies on further improvements in early-FTQC hardware or optimizing the time evolution circuits, potentially by using less exact approximations than trotterization.

\subsubsection{One-norm and qubitization}
In this section, we calculate the one-norm of the Hamiltonian, which in conjunction with the number of Pauli strings in the LCU allows us to get a schematic understanding of how these mappings would perform for a block encoding. A full resource estimation of an actual block encoding circuit is beyond the scope of this paper. However, it is known that the number of Toffoli gates required for the PREPARE and SELECT subroutines in Qubitization scales proportionally to $\lambda \cdot \sqrt{D}$ \cite{Berry2019}, where $D$ is the number of terms in the Hamiltonian and $\lambda = ||H||_{1}=\sum_{l}|w_l|$ is the one-norm for a Hamiltonian written as $H=\sum_{l}w_l P_l$, where $P_l$ are Pauli strings. We note that if symmetric terms are algorithmically grouped, this overhead can potentially be reduced to scale with $\sqrt{D_\text{unique}}$ where $D_\text{unique}$ is the number of unique coefficients. While exploiting this symmetry would lower the absolute cost for all mappings, we do not expect it to significantly alter the relative performance hierarchy between them. In Fig. \ref{fig:Onenormcomparison}, $\lambda $ and $\lambda \cdot \sqrt{D}$ are plotted for both the BHM and the HO models.

An interesting feature emerges when comparing the one-norms of the first-quantized mappings. Despite the B1Q mapping often generating a larger number of raw Pauli strings than the U1Q mapping as we saw in the last section, its one-norm $\lambda$ is suppressed and grows more slowly with system size. For example, in the HO model at $N=6,M=32$, the U1Q one-norm is  $4.0471 \cdot 10^4$, while the B1Q one-norm is only  $2.3254 \cdot 10^3$. 

We can understand this drastic reduction based on the triangle inequality. In the U1Q mapping, every matrix element is encoded to a distinct set of Pauli strings acting on dedicated qubits, meaning their coefficients sum strictly as absolute values. In contrast, the B1Q mapping encodes the physical Hilbert space into a logarithmic number of qubits. The required operator mapping leads to a large number of terms, but also means that multiple distinct one- and two-body physical terms can contribute the same Pauli strings. Because these elements naturally carry alternating signs, their coefficients can destructively interfere and algebraically cancel before the absolute value is taken ($\sum_i |c_i| \geq |\sum_i c_i|$). 

When evaluating the overall qubitization cost proxy $\lambda \cdot \sqrt{D}$, the B1Q mapping consistently outperforms the U1Q mapping as the system size grows. The destructive interference suppressing the one-norm is so significant that it outweighs the $\mathcal{O}(\sqrt{D})$ penalty incurred by the higher term count. This advantage becomes particularly pronounced at the structural sweet spots, identified in the last section, where $M=2^n$. Here all bit values are fully represented, leading to maximal algebraic cancellation that simultaneously minimizes both the one-norm and the total term count $D$. However, even for $M \neq 2^n$, the B1Q mapping remains the most efficient block-encoding choice at scale.

Across all measured system sizes, the second-quantized mappings (U2Q and B2Q) yield one-norms and values of $\lambda \cdot \sqrt{D}$ that are tens to hundreds of times larger than their first-quantized equivalents. This renders the second-quantized mappings practically uncompetitive for qubitization in models that conserve the number of particles.

\section{Discussion and outlook}
\label{sec:disussion}
In general, our results suggest that first quantized mappings are more gate-efficient at expressing the bosonic problem than the second quantized ones, when using the same basis functions. Naively, this can be understood from the fact that a factor of $(4 N)^k$ more Pauli strings are required to express off-diagonal $k$-RDM elements for the latter. The numerical results show that this indeed manifests in all the resource comparisons, and it is only for very small numbers of particles $N$ that the second quantized mappings can be more gate-efficient.  

For an accurate description of continuum problems, the number of modes $M$ needs to be much larger than the number of particles ($M \gg N$), so the binary first quantized mapping ($N \cdot \ceil{\log_2 (M)}$ qubits) is particularly useful for reducing the number of required qubits. For example, the HO for 6 particles in 128 modes (beyond what is solvable using exact diagonalization) can be implemented using $\ceil{\log_2(128)}\cdot 6 = 42$ logical qubits. However, it is possible for the binary second quantized mapping ($M \cdot \ceil{\log_2 (N+1)}$ qubits) to be more qubit-efficient, for example for investigating the Hubbard model at higher integer fillings where $N > M$. This will be at the cost of being highly gate-inefficient, as this use-case assumes large $N$ to go beyond classically solvable models.

In addition to the binary first quantized mapping, we investigated a unary first quantized mapping. This allowed us to obtain simple analytic formulas to compare the first and second quantized mappings, and we have shown that the unary first quantized mapping is in general the most gate-efficient. While this mapping is less qubit-efficient than the binary first and second quantized ones, requiring $N \cdot M$ qubits, it is comparable to the unary second quantized mapping ($M \cdot (N+1)$ qubits). It  typically uses fewer gates than the binary first quantized mapping for expressing the same operator, making it the best choice if gate-efficiency is more important than qubit-efficiency, although for complicated LCUs the binary mapping can be better for specific number of sites/modes, such as $M = 2^n$. Additionally we showed numerically that the binary first quantized mapping can be more efficient when considering a full simulation algorithm for the BHM and HO Hamiltonians. Looking beyond the simple gate counts of the LCU or a single Trotter step the binary mapping lowers the overall number of gates required for full qubitization-based QPE as it significantly lowers the one-norm of the Hamiltonian due to destructive interference between Pauli-strings. Investigating both binary and unary mappings for a Hamiltonian and algorithm of interest is therefore useful, as the former can potentially have a better gate-efficiency for a given algorithm using a much smaller number of qubits.

The unary mapping has one extra appealing feature: it results in a Hamiltonian which has the same form as the second quantized fermionic Hamiltonian in the Jordan-Wigner mapping (without enforcing anti-symmetry) and many previously proposed algorithms in the context of quantum chemistry are usable with only minor modifications.

In contrast with the fermionic case for which the second quantized mapping is generally more gate-efficient compared to the first quantized one for the same basis functions, our analytic and numeric results show that the opposite holds true for bosons. This suggests that first quantized mappings for bosonic problems are superior to second quantized mappings for both NISQ, early-FTQC, and FTQC devices in systems that \textit{conserve the number of bosons}. For problems where the number of bosons is not conserved first quantized mappings are inapplicable and the second quantized mappings are required. The main challenge for first quantized bosonic simulations is the preparation of a state with bosonic symmetry. As outlined in the introduction, there are already a large number of algorithms, such as QPE \cite{kitaev1995,Low2019,Kivlichan2020,Yuan2021} , TE-QSCI \cite{mikkelsen2024,sugisaki2024,yu2025}, and UCCSD-anzats based VQE \cite{Cederbaum2006,Tilly2021,Anand2022} that can be carried out based on initial input states where all bosons are in a single site/mode or where no site/mode has more than a single boson occupation, both of which can be straightforwardly prepared (the latter can be prepared by the method outlined in \cite{Berry2018}). However, finding an efficient way to symmetrize any input state is an important future research endeavor. 

Based on the estimates in this manuscript, the BHM is a good candidate for pushing system-sizes in the early-FTQC era as the number of $R_z(\phi)$ gates in the first quantized mapping is relatively small. Simulating the (long-term) dynamics of such models is classically difficult and therefore a promising candidate for quantum simulation.

\begin{acknowledgments}
The authors would like to thank Yuya Nakagawa for useful comments and discussion leading to major revisions of the manuscript. 
\end{acknowledgments}

\bibliographystyle{apsrev4-2}
\bibliography{library}

\begin{thebibliography}{53}%
\makeatletter
\providecommand \@ifxundefined [1]{%
 \@ifx{#1\undefined}
}%
\providecommand \@ifnum [1]{%
 \ifnum #1\expandafter \@firstoftwo
 \else \expandafter \@secondoftwo
 \fi
}%
\providecommand \@ifx [1]{%
 \ifx #1\expandafter \@firstoftwo
 \else \expandafter \@secondoftwo
 \fi
}%
\providecommand \natexlab [1]{#1}%
\providecommand \enquote  [1]{``#1''}%
\providecommand \bibnamefont  [1]{#1}%
\providecommand \bibfnamefont [1]{#1}%
\providecommand \citenamefont [1]{#1}%
\providecommand \href@noop [0]{\@secondoftwo}%
\providecommand \href [0]{\begingroup \@sanitize@url \@href}%
\providecommand \@href[1]{\@@startlink{#1}\@@href}%
\providecommand \@@href[1]{\endgroup#1\@@endlink}%
\providecommand \@sanitize@url [0]{\catcode `\\12\catcode `\$12\catcode
  `\&12\catcode `\#12\catcode `\^12\catcode `\_12\catcode `\%12\relax}%
\providecommand \@@startlink[1]{}%
\providecommand \@@endlink[0]{}%
\providecommand \url  [0]{\begingroup\@sanitize@url \@url }%
\providecommand \@url [1]{\endgroup\@href {#1}{\urlprefix }}%
\providecommand \urlprefix  [0]{URL }%
\providecommand \Eprint [0]{\href }%
\providecommand \doibase [0]{https://doi.org/}%
\providecommand \selectlanguage [0]{\@gobble}%
\providecommand \bibinfo  [0]{\@secondoftwo}%
\providecommand \bibfield  [0]{\@secondoftwo}%
\providecommand \translation [1]{[#1]}%
\providecommand \BibitemOpen [0]{}%
\providecommand \bibitemStop [0]{}%
\providecommand \bibitemNoStop [0]{.\EOS\space}%
\providecommand \EOS [0]{\spacefactor3000\relax}%
\providecommand \BibitemShut  [1]{\csname bibitem#1\endcsname}%
\let\auto@bib@innerbib\@empty
\bibitem [{\citenamefont {Feynman}(1982)}]{Feynman1982}%
  \BibitemOpen
  \bibfield  {author} {\bibinfo {author} {\bibfnamefont {R.~P.}\ \bibnamefont
  {Feynman}},\ }\href {https://doi.org/10.1007/BF02650179} {\bibfield
  {journal} {\bibinfo  {journal} {International Journal of Theoretical
  Physics}\ }\textbf {\bibinfo {volume} {21}},\ \bibinfo {pages} {467}
  (\bibinfo {year} {1982})}\BibitemShut {NoStop}%
\bibitem [{\citenamefont {Bloch}\ \emph {et~al.}(2008)\citenamefont {Bloch},
  \citenamefont {Dalibard},\ and\ \citenamefont {Zwerger}}]{Bloch2008}%
  \BibitemOpen
  \bibfield  {author} {\bibinfo {author} {\bibfnamefont {I.}~\bibnamefont
  {Bloch}}, \bibinfo {author} {\bibfnamefont {J.}~\bibnamefont {Dalibard}},\
  and\ \bibinfo {author} {\bibfnamefont {W.}~\bibnamefont {Zwerger}},\ }\href
  {https://doi.org/10.1103/RevModPhys.80.885} {\bibfield  {journal} {\bibinfo
  {journal} {Rev. Mod. Phys.}\ }\textbf {\bibinfo {volume} {80}},\ \bibinfo
  {pages} {885} (\bibinfo {year} {2008})}\BibitemShut {NoStop}%
\bibitem [{\citenamefont {Cazalilla}\ \emph {et~al.}(2011)\citenamefont
  {Cazalilla}, \citenamefont {Citro}, \citenamefont {Giamarchi}, \citenamefont
  {Orignac},\ and\ \citenamefont {Rigol}}]{Cazalilla2011}%
  \BibitemOpen
  \bibfield  {author} {\bibinfo {author} {\bibfnamefont {M.~A.}\ \bibnamefont
  {Cazalilla}}, \bibinfo {author} {\bibfnamefont {R.}~\bibnamefont {Citro}},
  \bibinfo {author} {\bibfnamefont {T.}~\bibnamefont {Giamarchi}}, \bibinfo
  {author} {\bibfnamefont {E.}~\bibnamefont {Orignac}},\ and\ \bibinfo {author}
  {\bibfnamefont {M.}~\bibnamefont {Rigol}},\ }\href
  {https://doi.org/10.1103/RevModPhys.83.1405} {\bibfield  {journal} {\bibinfo
  {journal} {Rev. Mod. Phys.}\ }\textbf {\bibinfo {volume} {83}},\ \bibinfo
  {pages} {1405} (\bibinfo {year} {2011})}\BibitemShut {NoStop}%
\bibitem [{\citenamefont {Z\"urn}\ \emph {et~al.}(2012)\citenamefont {Z\"urn},
  \citenamefont {Serwane}, \citenamefont {Lompe}, \citenamefont {Wenz},
  \citenamefont {Ries}, \citenamefont {Bohn},\ and\ \citenamefont
  {Jochim}}]{Jochim2012}%
  \BibitemOpen
  \bibfield  {author} {\bibinfo {author} {\bibfnamefont {G.}~\bibnamefont
  {Z\"urn}}, \bibinfo {author} {\bibfnamefont {F.}~\bibnamefont {Serwane}},
  \bibinfo {author} {\bibfnamefont {T.}~\bibnamefont {Lompe}}, \bibinfo
  {author} {\bibfnamefont {A.~N.}\ \bibnamefont {Wenz}}, \bibinfo {author}
  {\bibfnamefont {M.~G.}\ \bibnamefont {Ries}}, \bibinfo {author}
  {\bibfnamefont {J.~E.}\ \bibnamefont {Bohn}},\ and\ \bibinfo {author}
  {\bibfnamefont {S.}~\bibnamefont {Jochim}},\ }\href
  {https://doi.org/10.1103/PhysRevLett.108.075303} {\bibfield  {journal}
  {\bibinfo  {journal} {Phys. Rev. Lett.}\ }\textbf {\bibinfo {volume} {108}},\
  \bibinfo {pages} {075303} (\bibinfo {year} {2012})}\BibitemShut {NoStop}%
\bibitem [{\citenamefont {Mistakidis}\ \emph {et~al.}(2023)\citenamefont
  {Mistakidis}, \citenamefont {Volosniev}, \citenamefont {Barfknecht},
  \citenamefont {Fogarty}, \citenamefont {Busch}, \citenamefont {Foerster},
  \citenamefont {Schmelcher},\ and\ \citenamefont {Zinner}}]{Mistakidis2023}%
  \BibitemOpen
  \bibfield  {author} {\bibinfo {author} {\bibfnamefont {S.}~\bibnamefont
  {Mistakidis}}, \bibinfo {author} {\bibfnamefont {A.}~\bibnamefont
  {Volosniev}}, \bibinfo {author} {\bibfnamefont {R.}~\bibnamefont
  {Barfknecht}}, \bibinfo {author} {\bibfnamefont {T.}~\bibnamefont {Fogarty}},
  \bibinfo {author} {\bibfnamefont {T.}~\bibnamefont {Busch}}, \bibinfo
  {author} {\bibfnamefont {A.}~\bibnamefont {Foerster}}, \bibinfo {author}
  {\bibfnamefont {P.}~\bibnamefont {Schmelcher}},\ and\ \bibinfo {author}
  {\bibfnamefont {N.}~\bibnamefont {Zinner}},\ }\href
  {https://doi.org/https://doi.org/10.1016/j.physrep.2023.10.004} {\bibfield
  {journal} {\bibinfo  {journal} {Physics Reports}\ }\textbf {\bibinfo {volume}
  {1042}},\ \bibinfo {pages} {1} (\bibinfo {year} {2023})},\ \bibinfo {note}
  {few-body Bose gases in low dimensions—A laboratory for quantum
  dynamics}\BibitemShut {NoStop}%
\bibitem [{\citenamefont {Nielsen}\ and\ \citenamefont
  {Chuang}(2010)}]{Nielsen_Chuang2010}%
  \BibitemOpen
  \bibfield  {author} {\bibinfo {author} {\bibfnamefont {M.~A.}\ \bibnamefont
  {Nielsen}}\ and\ \bibinfo {author} {\bibfnamefont {I.~L.}\ \bibnamefont
  {Chuang}},\ }\href@noop {} {\emph {\bibinfo {title} {Quantum Computation and
  Quantum Information: 10th Anniversary Edition}}}\ (\bibinfo  {publisher}
  {Cambridge University Press},\ \bibinfo {year} {2010})\BibitemShut {NoStop}%
\bibitem [{\citenamefont {de~Leon}\ \emph {et~al.}(2021)\citenamefont
  {de~Leon}, \citenamefont {Itoh}, \citenamefont {Kim}, \citenamefont {Mehta},
  \citenamefont {Northup}, \citenamefont {Paik}, \citenamefont {Palmer},
  \citenamefont {Samarth}, \citenamefont {Sangtawesin},\ and\ \citenamefont
  {Steuerman}}]{Leon2021}%
  \BibitemOpen
  \bibfield  {author} {\bibinfo {author} {\bibfnamefont {N.~P.}\ \bibnamefont
  {de~Leon}}, \bibinfo {author} {\bibfnamefont {K.~M.}\ \bibnamefont {Itoh}},
  \bibinfo {author} {\bibfnamefont {D.}~\bibnamefont {Kim}}, \bibinfo {author}
  {\bibfnamefont {K.~K.}\ \bibnamefont {Mehta}}, \bibinfo {author}
  {\bibfnamefont {T.~E.}\ \bibnamefont {Northup}}, \bibinfo {author}
  {\bibfnamefont {H.}~\bibnamefont {Paik}}, \bibinfo {author} {\bibfnamefont
  {B.~S.}\ \bibnamefont {Palmer}}, \bibinfo {author} {\bibfnamefont
  {N.}~\bibnamefont {Samarth}}, \bibinfo {author} {\bibfnamefont
  {S.}~\bibnamefont {Sangtawesin}},\ and\ \bibinfo {author} {\bibfnamefont
  {D.~W.}\ \bibnamefont {Steuerman}},\ }\href
  {https://doi.org/10.1126/science.abb2823} {\bibfield  {journal} {\bibinfo
  {journal} {Science}\ }\textbf {\bibinfo {volume} {372}},\ \bibinfo {pages}
  {eabb2823} (\bibinfo {year} {2021})},\ \Eprint
  {https://arxiv.org/abs/https://www.science.org/doi/pdf/10.1126/science.abb2823}
  {https://www.science.org/doi/pdf/10.1126/science.abb2823} \BibitemShut
  {NoStop}%
\bibitem [{\citenamefont {Wang}\ \emph {et~al.}(2020)\citenamefont {Wang},
  \citenamefont {Hu}, \citenamefont {Sanders},\ and\ \citenamefont
  {Kais}}]{Wang2020}%
  \BibitemOpen
  \bibfield  {author} {\bibinfo {author} {\bibfnamefont {Y.}~\bibnamefont
  {Wang}}, \bibinfo {author} {\bibfnamefont {Z.}~\bibnamefont {Hu}}, \bibinfo
  {author} {\bibfnamefont {B.~C.}\ \bibnamefont {Sanders}},\ and\ \bibinfo
  {author} {\bibfnamefont {S.}~\bibnamefont {Kais}},\ }\bibfield  {journal}
  {\bibinfo  {journal} {Frontiers in Physics}\ }\textbf {\bibinfo {volume}
  {8}},\ \href {https://doi.org/10.3389/fphy.2020.589504}
  {10.3389/fphy.2020.589504} (\bibinfo {year} {2020})\BibitemShut {NoStop}%
\bibitem [{\citenamefont {Ringbauer}\ \emph {et~al.}(2022)\citenamefont
  {Ringbauer}, \citenamefont {Meth}, \citenamefont {Postler}, \citenamefont
  {Stricker}, \citenamefont {Blatt}, \citenamefont {Schindler},\ and\
  \citenamefont {Monz}}]{Ringbauer2022}%
  \BibitemOpen
  \bibfield  {author} {\bibinfo {author} {\bibfnamefont {M.}~\bibnamefont
  {Ringbauer}}, \bibinfo {author} {\bibfnamefont {M.}~\bibnamefont {Meth}},
  \bibinfo {author} {\bibfnamefont {L.}~\bibnamefont {Postler}}, \bibinfo
  {author} {\bibfnamefont {R.}~\bibnamefont {Stricker}}, \bibinfo {author}
  {\bibfnamefont {R.}~\bibnamefont {Blatt}}, \bibinfo {author} {\bibfnamefont
  {P.}~\bibnamefont {Schindler}},\ and\ \bibinfo {author} {\bibfnamefont
  {T.}~\bibnamefont {Monz}},\ }\href
  {https://doi.org/10.1038/s41567-022-01658-0} {\bibfield  {journal} {\bibinfo
  {journal} {Nature Physics}\ }\textbf {\bibinfo {volume} {18}},\ \bibinfo
  {pages} {1053} (\bibinfo {year} {2022})}\BibitemShut {NoStop}%
\bibitem [{\citenamefont {Cao}\ \emph {et~al.}(2019)\citenamefont {Cao},
  \citenamefont {Romero}, \citenamefont {Olson}, \citenamefont {Degroote},
  \citenamefont {Johnson}, \citenamefont {Kieferov{\'a}}, \citenamefont
  {Kivlichan}, \citenamefont {Menke}, \citenamefont {Peropadre}, \citenamefont
  {Sawaya}, \citenamefont {Sim}, \citenamefont {Veis},\ and\ \citenamefont
  {Aspuru-Guzik}}]{Cao2019}%
  \BibitemOpen
  \bibfield  {author} {\bibinfo {author} {\bibfnamefont {Y.}~\bibnamefont
  {Cao}}, \bibinfo {author} {\bibfnamefont {J.}~\bibnamefont {Romero}},
  \bibinfo {author} {\bibfnamefont {J.~P.}\ \bibnamefont {Olson}}, \bibinfo
  {author} {\bibfnamefont {M.}~\bibnamefont {Degroote}}, \bibinfo {author}
  {\bibfnamefont {P.~D.}\ \bibnamefont {Johnson}}, \bibinfo {author}
  {\bibfnamefont {M.}~\bibnamefont {Kieferov{\'a}}}, \bibinfo {author}
  {\bibfnamefont {I.~D.}\ \bibnamefont {Kivlichan}}, \bibinfo {author}
  {\bibfnamefont {T.}~\bibnamefont {Menke}}, \bibinfo {author} {\bibfnamefont
  {B.}~\bibnamefont {Peropadre}}, \bibinfo {author} {\bibfnamefont {N.~P.~D.}\
  \bibnamefont {Sawaya}}, \bibinfo {author} {\bibfnamefont {S.}~\bibnamefont
  {Sim}}, \bibinfo {author} {\bibfnamefont {L.}~\bibnamefont {Veis}},\ and\
  \bibinfo {author} {\bibfnamefont {A.}~\bibnamefont {Aspuru-Guzik}},\ }\href
  {https://doi.org/10.1021/acs.chemrev.8b00803} {\bibfield  {journal} {\bibinfo
   {journal} {Chemical Reviews}\ }\textbf {\bibinfo {volume} {119}},\ \bibinfo
  {pages} {10856} (\bibinfo {year} {2019})}\BibitemShut {NoStop}%
\bibitem [{\citenamefont {McArdle}\ \emph {et~al.}(2020)\citenamefont
  {McArdle}, \citenamefont {Endo}, \citenamefont {Aspuru-Guzik}, \citenamefont
  {Benjamin},\ and\ \citenamefont {Yuan}}]{McArdle2020}%
  \BibitemOpen
  \bibfield  {author} {\bibinfo {author} {\bibfnamefont {S.}~\bibnamefont
  {McArdle}}, \bibinfo {author} {\bibfnamefont {S.}~\bibnamefont {Endo}},
  \bibinfo {author} {\bibfnamefont {A.}~\bibnamefont {Aspuru-Guzik}}, \bibinfo
  {author} {\bibfnamefont {S.~C.}\ \bibnamefont {Benjamin}},\ and\ \bibinfo
  {author} {\bibfnamefont {X.}~\bibnamefont {Yuan}},\ }\href
  {https://doi.org/10.1103/RevModPhys.92.015003} {\bibfield  {journal}
  {\bibinfo  {journal} {Rev. Mod. Phys.}\ }\textbf {\bibinfo {volume} {92}},\
  \bibinfo {pages} {015003} (\bibinfo {year} {2020})}\BibitemShut {NoStop}%
\bibitem [{\citenamefont {Somma}\ \emph {et~al.}(2003)\citenamefont {Somma},
  \citenamefont {Ortiz}, \citenamefont {Knill},\ and\ \citenamefont
  {Gubernatis}}]{Rolando2003}%
  \BibitemOpen
  \bibfield  {author} {\bibinfo {author} {\bibfnamefont {R.~D.}\ \bibnamefont
  {Somma}}, \bibinfo {author} {\bibfnamefont {G.}~\bibnamefont {Ortiz}},
  \bibinfo {author} {\bibfnamefont {E.~H.}\ \bibnamefont {Knill}},\ and\
  \bibinfo {author} {\bibfnamefont {J.}~\bibnamefont {Gubernatis}},\ }in\ \href
  {https://doi.org/10.1117/12.487249} {\emph {\bibinfo {booktitle} {Quantum
  Information and Computation}}},\ Vol.\ \bibinfo {volume} {5105},\ \bibinfo
  {editor} {edited by\ \bibinfo {editor} {\bibfnamefont {E.}~\bibnamefont
  {Donkor}}, \bibinfo {editor} {\bibfnamefont {A.~R.}\ \bibnamefont {Pirich}},\
  and\ \bibinfo {editor} {\bibfnamefont {H.~E.}\ \bibnamefont {Brandt}}},\
  \bibinfo {organization} {International Society for Optics and Photonics}\
  (\bibinfo  {publisher} {SPIE},\ \bibinfo {year} {2003})\ pp.\ \bibinfo
  {pages} {96 -- 103}\BibitemShut {NoStop}%
\bibitem [{\citenamefont {Sawaya}\ \emph
  {et~al.}(2020{\natexlab{a}})\citenamefont {Sawaya}, \citenamefont {Menke},
  \citenamefont {Kyaw}, \citenamefont {Johri}, \citenamefont {Aspuru-Guzik},\
  and\ \citenamefont {Guerreschi}}]{Sawaya2020a}%
  \BibitemOpen
  \bibfield  {author} {\bibinfo {author} {\bibfnamefont {N.~P.~D.}\
  \bibnamefont {Sawaya}}, \bibinfo {author} {\bibfnamefont {T.}~\bibnamefont
  {Menke}}, \bibinfo {author} {\bibfnamefont {T.~H.}\ \bibnamefont {Kyaw}},
  \bibinfo {author} {\bibfnamefont {S.}~\bibnamefont {Johri}}, \bibinfo
  {author} {\bibfnamefont {A.}~\bibnamefont {Aspuru-Guzik}},\ and\ \bibinfo
  {author} {\bibfnamefont {G.~G.}\ \bibnamefont {Guerreschi}},\ }\href
  {https://doi.org/10.1038/s41534-020-0278-0} {\bibfield  {journal} {\bibinfo
  {journal} {npj Quantum Information}\ }\textbf {\bibinfo {volume} {6}},\
  \bibinfo {pages} {49} (\bibinfo {year} {2020}{\natexlab{a}})}\BibitemShut
  {NoStop}%
\bibitem [{\citenamefont {Trenev}\ \emph {et~al.}(2023)\citenamefont {Trenev},
  \citenamefont {Ollitrault}, \citenamefont {Harwood}, \citenamefont
  {Gujarati}, \citenamefont {Raman}, \citenamefont {Mezzacapo},\ and\
  \citenamefont {Mostame}}]{trenev2023}%
  \BibitemOpen
  \bibfield  {author} {\bibinfo {author} {\bibfnamefont {D.}~\bibnamefont
  {Trenev}}, \bibinfo {author} {\bibfnamefont {P.~J.}\ \bibnamefont
  {Ollitrault}}, \bibinfo {author} {\bibfnamefont {S.~M.}\ \bibnamefont
  {Harwood}}, \bibinfo {author} {\bibfnamefont {T.~P.}\ \bibnamefont
  {Gujarati}}, \bibinfo {author} {\bibfnamefont {S.}~\bibnamefont {Raman}},
  \bibinfo {author} {\bibfnamefont {A.}~\bibnamefont {Mezzacapo}},\ and\
  \bibinfo {author} {\bibfnamefont {S.}~\bibnamefont {Mostame}},\ }\href@noop
  {} {\bibinfo {title} {Refining resource estimation for the quantum
  computation of vibrational molecular spectra through trotter error analysis}}
  (\bibinfo {year} {2023}),\ \Eprint {https://arxiv.org/abs/2311.03719}
  {arXiv:2311.03719 [quant-ph]} \BibitemShut {NoStop}%
\bibitem [{\citenamefont {Majland}\ and\ \citenamefont
  {Zinner}(2021)}]{majland2021}%
  \BibitemOpen
  \bibfield  {author} {\bibinfo {author} {\bibfnamefont {M.}~\bibnamefont
  {Majland}}\ and\ \bibinfo {author} {\bibfnamefont {N.~T.}\ \bibnamefont
  {Zinner}},\ }\href@noop {} {\bibinfo {title} {Resource-efficient encoding
  algorithm for variational bosonic quantum simulations}} (\bibinfo {year}
  {2021}),\ \Eprint {https://arxiv.org/abs/2102.11886} {arXiv:2102.11886
  [quant-ph]} \BibitemShut {NoStop}%
\bibitem [{\citenamefont {Kuwahara}\ \emph {et~al.}(2024)\citenamefont
  {Kuwahara}, \citenamefont {Vu},\ and\ \citenamefont {Saito}}]{Kuwahara2024}%
  \BibitemOpen
  \bibfield  {author} {\bibinfo {author} {\bibfnamefont {T.}~\bibnamefont
  {Kuwahara}}, \bibinfo {author} {\bibfnamefont {T.~V.}\ \bibnamefont {Vu}},\
  and\ \bibinfo {author} {\bibfnamefont {K.}~\bibnamefont {Saito}},\ }\href
  {https://doi.org/10.1038/s41467-024-46501-7} {\bibfield  {journal} {\bibinfo
  {journal} {Nature Communications}\ }\textbf {\bibinfo {volume} {15}},\
  \bibinfo {pages} {2520} (\bibinfo {year} {2024})}\BibitemShut {NoStop}%
\bibitem [{\citenamefont {Peng}\ \emph {et~al.}(2023)\citenamefont {Peng},
  \citenamefont {Su}, \citenamefont {Claudino}, \citenamefont {Kowalski},
  \citenamefont {Low},\ and\ \citenamefont {Roetteler}}]{peng2023}%
  \BibitemOpen
  \bibfield  {author} {\bibinfo {author} {\bibfnamefont {B.}~\bibnamefont
  {Peng}}, \bibinfo {author} {\bibfnamefont {Y.}~\bibnamefont {Su}}, \bibinfo
  {author} {\bibfnamefont {D.}~\bibnamefont {Claudino}}, \bibinfo {author}
  {\bibfnamefont {K.}~\bibnamefont {Kowalski}}, \bibinfo {author}
  {\bibfnamefont {G.~H.}\ \bibnamefont {Low}},\ and\ \bibinfo {author}
  {\bibfnamefont {M.}~\bibnamefont {Roetteler}},\ }\href@noop {} {\bibinfo
  {title} {Quantum simulation of boson-related hamiltonians: Techniques,
  effective hamiltonian construction, and error analysis}} (\bibinfo {year}
  {2023}),\ \Eprint {https://arxiv.org/abs/2307.06580} {arXiv:2307.06580
  [quant-ph]} \BibitemShut {NoStop}%
\bibitem [{\citenamefont {McArdle}\ \emph {et~al.}(2019)\citenamefont
  {McArdle}, \citenamefont {Mayorov}, \citenamefont {Shan}, \citenamefont
  {Benjamin},\ and\ \citenamefont {Yuan}}]{McArdle2019}%
  \BibitemOpen
  \bibfield  {author} {\bibinfo {author} {\bibfnamefont {S.}~\bibnamefont
  {McArdle}}, \bibinfo {author} {\bibfnamefont {A.}~\bibnamefont {Mayorov}},
  \bibinfo {author} {\bibfnamefont {X.}~\bibnamefont {Shan}}, \bibinfo {author}
  {\bibfnamefont {S.}~\bibnamefont {Benjamin}},\ and\ \bibinfo {author}
  {\bibfnamefont {X.}~\bibnamefont {Yuan}},\ }\href
  {https://doi.org/10.1039/C9SC01313J} {\bibfield  {journal} {\bibinfo
  {journal} {Chem. Sci.}\ }\textbf {\bibinfo {volume} {10}},\ \bibinfo {pages}
  {5725} (\bibinfo {year} {2019})}\BibitemShut {NoStop}%
\bibitem [{\citenamefont {Sawaya}\ and\ \citenamefont
  {Huh}(2019)}]{Sawaya2019}%
  \BibitemOpen
  \bibfield  {author} {\bibinfo {author} {\bibfnamefont {N.~P.~D.}\
  \bibnamefont {Sawaya}}\ and\ \bibinfo {author} {\bibfnamefont
  {J.}~\bibnamefont {Huh}},\ }\href
  {https://doi.org/10.1021/acs.jpclett.9b01117} {\bibfield  {journal} {\bibinfo
   {journal} {The Journal of Physical Chemistry Letters}\ }\textbf {\bibinfo
  {volume} {10}},\ \bibinfo {pages} {3586} (\bibinfo {year}
  {2019})}\BibitemShut {NoStop}%
\bibitem [{\citenamefont {L\"otstedt}\ \emph {et~al.}(2021)\citenamefont
  {L\"otstedt}, \citenamefont {Yamanouchi}, \citenamefont {Tsuchiya},\ and\
  \citenamefont {Tachikawa}}]{Lotstedt2021}%
  \BibitemOpen
  \bibfield  {author} {\bibinfo {author} {\bibfnamefont {E.}~\bibnamefont
  {L\"otstedt}}, \bibinfo {author} {\bibfnamefont {K.}~\bibnamefont
  {Yamanouchi}}, \bibinfo {author} {\bibfnamefont {T.}~\bibnamefont
  {Tsuchiya}},\ and\ \bibinfo {author} {\bibfnamefont {Y.}~\bibnamefont
  {Tachikawa}},\ }\href {https://doi.org/10.1103/PhysRevA.103.062609}
  {\bibfield  {journal} {\bibinfo  {journal} {Phys. Rev. A}\ }\textbf {\bibinfo
  {volume} {103}},\ \bibinfo {pages} {062609} (\bibinfo {year}
  {2021})}\BibitemShut {NoStop}%
\bibitem [{\citenamefont {Sawaya}\ \emph {et~al.}(2021)\citenamefont {Sawaya},
  \citenamefont {Paesani},\ and\ \citenamefont {Tabor}}]{Sawaya2021}%
  \BibitemOpen
  \bibfield  {author} {\bibinfo {author} {\bibfnamefont {N.~P.~D.}\
  \bibnamefont {Sawaya}}, \bibinfo {author} {\bibfnamefont {F.}~\bibnamefont
  {Paesani}},\ and\ \bibinfo {author} {\bibfnamefont {D.~P.}\ \bibnamefont
  {Tabor}},\ }\href {https://doi.org/10.1103/PhysRevA.104.062419} {\bibfield
  {journal} {\bibinfo  {journal} {Phys. Rev. A}\ }\textbf {\bibinfo {volume}
  {104}},\ \bibinfo {pages} {062419} (\bibinfo {year} {2021})}\BibitemShut
  {NoStop}%
\bibitem [{\citenamefont {Sawaya}\ \emph
  {et~al.}(2020{\natexlab{b}})\citenamefont {Sawaya}, \citenamefont
  {Guerreschi},\ and\ \citenamefont {Holmes}}]{Sawaya2020b}%
  \BibitemOpen
  \bibfield  {author} {\bibinfo {author} {\bibfnamefont {N.~P.~D.}\
  \bibnamefont {Sawaya}}, \bibinfo {author} {\bibfnamefont {G.~G.}\
  \bibnamefont {Guerreschi}},\ and\ \bibinfo {author} {\bibfnamefont
  {A.}~\bibnamefont {Holmes}},\ }in\ \href
  {https://doi.org/10.1109/QCE49297.2020.00031} {\emph {\bibinfo {booktitle}
  {2020 IEEE International Conference on Quantum Computing and Engineering
  (QCE)}}}\ (\bibinfo {year} {2020})\ pp.\ \bibinfo {pages}
  {180--190}\BibitemShut {NoStop}%
\bibitem [{\citenamefont {Bahrami}\ and\ \citenamefont
  {Sawaya}(2024)}]{bahrami2024}%
  \BibitemOpen
  \bibfield  {author} {\bibinfo {author} {\bibfnamefont {S.}~\bibnamefont
  {Bahrami}}\ and\ \bibinfo {author} {\bibfnamefont {N.}~\bibnamefont
  {Sawaya}},\ }\href@noop {} {\bibinfo {title} {Particle-conserving quantum
  circuit ansatz with applications in variational simulation of bosonic
  systems}} (\bibinfo {year} {2024}),\ \Eprint
  {https://arxiv.org/abs/2402.18768} {arXiv:2402.18768 [quant-ph]} \BibitemShut
  {NoStop}%
\bibitem [{\citenamefont {Liu}\ \emph {et~al.}(2022)\citenamefont {Liu},
  \citenamefont {Li}, \citenamefont {Zheng}, \citenamefont {Yuan},\ and\
  \citenamefont {Sun}}]{Liu2022}%
  \BibitemOpen
  \bibfield  {author} {\bibinfo {author} {\bibfnamefont {J.}~\bibnamefont
  {Liu}}, \bibinfo {author} {\bibfnamefont {Z.}~\bibnamefont {Li}}, \bibinfo
  {author} {\bibfnamefont {H.}~\bibnamefont {Zheng}}, \bibinfo {author}
  {\bibfnamefont {X.}~\bibnamefont {Yuan}},\ and\ \bibinfo {author}
  {\bibfnamefont {J.}~\bibnamefont {Sun}},\ }\href
  {https://doi.org/10.1088/2632-2153/aca06b} {\bibfield  {journal} {\bibinfo
  {journal} {Machine Learning: Science and Technology}\ }\textbf {\bibinfo
  {volume} {3}},\ \bibinfo {pages} {045030} (\bibinfo {year}
  {2022})}\BibitemShut {NoStop}%
\bibitem [{\citenamefont {Tudorovskaya}\ and\ \citenamefont {Mu\~noz
  Ramo}(2024)}]{Tudorovskaya2024}%
  \BibitemOpen
  \bibfield  {author} {\bibinfo {author} {\bibfnamefont {M.}~\bibnamefont
  {Tudorovskaya}}\ and\ \bibinfo {author} {\bibfnamefont {D.}~\bibnamefont
  {Mu\~noz Ramo}},\ }\href {https://doi.org/10.1103/PhysRevA.109.032612}
  {\bibfield  {journal} {\bibinfo  {journal} {Phys. Rev. A}\ }\textbf {\bibinfo
  {volume} {109}},\ \bibinfo {pages} {032612} (\bibinfo {year}
  {2024})}\BibitemShut {NoStop}%
\bibitem [{\citenamefont {Wang}\ \emph {et~al.}(2023)\citenamefont {Wang},
  \citenamefont {Sager-Smith},\ and\ \citenamefont {Mazziotti}}]{Wang2023}%
  \BibitemOpen
  \bibfield  {author} {\bibinfo {author} {\bibfnamefont {Y.}~\bibnamefont
  {Wang}}, \bibinfo {author} {\bibfnamefont {L.~M.}\ \bibnamefont
  {Sager-Smith}},\ and\ \bibinfo {author} {\bibfnamefont {D.~A.}\ \bibnamefont
  {Mazziotti}},\ }\href {https://doi.org/10.1088/1367-2630/acf9c3} {\bibfield
  {journal} {\bibinfo  {journal} {New Journal of Physics}\ }\textbf {\bibinfo
  {volume} {25}},\ \bibinfo {pages} {103005} (\bibinfo {year}
  {2023})}\BibitemShut {NoStop}%
\bibitem [{\citenamefont {Abrams}\ and\ \citenamefont
  {Lloyd}(1997)}]{Abrams1997}%
  \BibitemOpen
  \bibfield  {author} {\bibinfo {author} {\bibfnamefont {D.~S.}\ \bibnamefont
  {Abrams}}\ and\ \bibinfo {author} {\bibfnamefont {S.}~\bibnamefont {Lloyd}},\
  }\href {https://doi.org/10.1103/PhysRevLett.79.2586} {\bibfield  {journal}
  {\bibinfo  {journal} {Phys. Rev. Lett.}\ }\textbf {\bibinfo {volume} {79}},\
  \bibinfo {pages} {2586} (\bibinfo {year} {1997})}\BibitemShut {NoStop}%
\bibitem [{\citenamefont {Su}\ \emph {et~al.}(2021)\citenamefont {Su},
  \citenamefont {Berry}, \citenamefont {Wiebe}, \citenamefont {Rubin},\ and\
  \citenamefont {Babbush}}]{Yuan2021}%
  \BibitemOpen
  \bibfield  {author} {\bibinfo {author} {\bibfnamefont {Y.}~\bibnamefont
  {Su}}, \bibinfo {author} {\bibfnamefont {D.~W.}\ \bibnamefont {Berry}},
  \bibinfo {author} {\bibfnamefont {N.}~\bibnamefont {Wiebe}}, \bibinfo
  {author} {\bibfnamefont {N.}~\bibnamefont {Rubin}},\ and\ \bibinfo {author}
  {\bibfnamefont {R.}~\bibnamefont {Babbush}},\ }\href
  {https://doi.org/10.1103/PRXQuantum.2.040332} {\bibfield  {journal} {\bibinfo
   {journal} {PRX Quantum}\ }\textbf {\bibinfo {volume} {2}},\ \bibinfo {pages}
  {040332} (\bibinfo {year} {2021})}\BibitemShut {NoStop}%
\bibitem [{\citenamefont {Delgado}\ \emph {et~al.}(2022)\citenamefont
  {Delgado}, \citenamefont {Casares}, \citenamefont {dos Reis}, \citenamefont
  {Zini}, \citenamefont {Campos}, \citenamefont {Cruz-Hern\'andez},
  \citenamefont {Voigt}, \citenamefont {Lowe}, \citenamefont {Jahangiri},
  \citenamefont {Martin-Delgado}, \citenamefont {Mueller},\ and\ \citenamefont
  {Arrazola}}]{Delgado2021}%
  \BibitemOpen
  \bibfield  {author} {\bibinfo {author} {\bibfnamefont {A.}~\bibnamefont
  {Delgado}}, \bibinfo {author} {\bibfnamefont {P.~A.~M.}\ \bibnamefont
  {Casares}}, \bibinfo {author} {\bibfnamefont {R.}~\bibnamefont {dos Reis}},
  \bibinfo {author} {\bibfnamefont {M.~S.}\ \bibnamefont {Zini}}, \bibinfo
  {author} {\bibfnamefont {R.}~\bibnamefont {Campos}}, \bibinfo {author}
  {\bibfnamefont {N.}~\bibnamefont {Cruz-Hern\'andez}}, \bibinfo {author}
  {\bibfnamefont {A.-C.}\ \bibnamefont {Voigt}}, \bibinfo {author}
  {\bibfnamefont {A.}~\bibnamefont {Lowe}}, \bibinfo {author} {\bibfnamefont
  {S.}~\bibnamefont {Jahangiri}}, \bibinfo {author} {\bibfnamefont {M.~A.}\
  \bibnamefont {Martin-Delgado}}, \bibinfo {author} {\bibfnamefont {J.~E.}\
  \bibnamefont {Mueller}},\ and\ \bibinfo {author} {\bibfnamefont {J.~M.}\
  \bibnamefont {Arrazola}},\ }\href
  {https://doi.org/10.1103/PhysRevA.106.032428} {\bibfield  {journal} {\bibinfo
   {journal} {Phys. Rev. A}\ }\textbf {\bibinfo {volume} {106}},\ \bibinfo
  {pages} {032428} (\bibinfo {year} {2022})}\BibitemShut {NoStop}%
\bibitem [{\citenamefont {Rubin}\ \emph {et~al.}(2023)\citenamefont {Rubin},
  \citenamefont {Berry}, \citenamefont {Malone}, \citenamefont {White},
  \citenamefont {Khattar}, \citenamefont {DePrince}, \citenamefont {Sicolo},
  \citenamefont {K\"uehn}, \citenamefont {Kaicher}, \citenamefont {Lee},\ and\
  \citenamefont {Babbush}}]{Rubin2023}%
  \BibitemOpen
  \bibfield  {author} {\bibinfo {author} {\bibfnamefont {N.~C.}\ \bibnamefont
  {Rubin}}, \bibinfo {author} {\bibfnamefont {D.~W.}\ \bibnamefont {Berry}},
  \bibinfo {author} {\bibfnamefont {F.~D.}\ \bibnamefont {Malone}}, \bibinfo
  {author} {\bibfnamefont {A.~F.}\ \bibnamefont {White}}, \bibinfo {author}
  {\bibfnamefont {T.}~\bibnamefont {Khattar}}, \bibinfo {author} {\bibfnamefont
  {A.~E.}\ \bibnamefont {DePrince}}, \bibinfo {author} {\bibfnamefont
  {S.}~\bibnamefont {Sicolo}}, \bibinfo {author} {\bibfnamefont
  {M.}~\bibnamefont {K\"uehn}}, \bibinfo {author} {\bibfnamefont
  {M.}~\bibnamefont {Kaicher}}, \bibinfo {author} {\bibfnamefont
  {J.}~\bibnamefont {Lee}},\ and\ \bibinfo {author} {\bibfnamefont
  {R.}~\bibnamefont {Babbush}},\ }\href
  {https://doi.org/10.1103/PRXQuantum.4.040303} {\bibfield  {journal} {\bibinfo
   {journal} {PRX Quantum}\ }\textbf {\bibinfo {volume} {4}},\ \bibinfo {pages}
  {040303} (\bibinfo {year} {2023})}\BibitemShut {NoStop}%
\bibitem [{\citenamefont {Shokrian~Zini}\ \emph {et~al.}(2023)\citenamefont
  {Shokrian~Zini}, \citenamefont {Delgado}, \citenamefont {dos Reis},
  \citenamefont {Moreno~Casares}, \citenamefont {Mueller}, \citenamefont
  {Voigt},\ and\ \citenamefont {Arrazola}}]{Shokrian2023}%
  \BibitemOpen
  \bibfield  {author} {\bibinfo {author} {\bibfnamefont {M.}~\bibnamefont
  {Shokrian~Zini}}, \bibinfo {author} {\bibfnamefont {A.}~\bibnamefont
  {Delgado}}, \bibinfo {author} {\bibfnamefont {R.}~\bibnamefont {dos Reis}},
  \bibinfo {author} {\bibfnamefont {P.~A.}\ \bibnamefont {Moreno~Casares}},
  \bibinfo {author} {\bibfnamefont {J.~E.}\ \bibnamefont {Mueller}}, \bibinfo
  {author} {\bibfnamefont {A.-C.}\ \bibnamefont {Voigt}},\ and\ \bibinfo
  {author} {\bibfnamefont {J.~M.}\ \bibnamefont {Arrazola}},\ }\href
  {https://doi.org/10.22331/q-2023-07-10-1049} {\bibfield  {journal} {\bibinfo
  {journal} {{Quantum}}\ }\textbf {\bibinfo {volume} {7}},\ \bibinfo {pages}
  {1049} (\bibinfo {year} {2023})}\BibitemShut {NoStop}%
\bibitem [{\citenamefont {Berry}\ \emph {et~al.}(2024)\citenamefont {Berry},
  \citenamefont {Rubin}, \citenamefont {Elnabawy}, \citenamefont {Ahlers},
  \citenamefont {III}, \citenamefont {Lee}, \citenamefont {Gogolin},\ and\
  \citenamefont {Babbush}}]{berry2024}%
  \BibitemOpen
  \bibfield  {author} {\bibinfo {author} {\bibfnamefont {D.~W.}\ \bibnamefont
  {Berry}}, \bibinfo {author} {\bibfnamefont {N.~C.}\ \bibnamefont {Rubin}},
  \bibinfo {author} {\bibfnamefont {A.~O.}\ \bibnamefont {Elnabawy}}, \bibinfo
  {author} {\bibfnamefont {G.}~\bibnamefont {Ahlers}}, \bibinfo {author}
  {\bibfnamefont {A.~E.~D.}\ \bibnamefont {III}}, \bibinfo {author}
  {\bibfnamefont {J.}~\bibnamefont {Lee}}, \bibinfo {author} {\bibfnamefont
  {C.}~\bibnamefont {Gogolin}},\ and\ \bibinfo {author} {\bibfnamefont
  {R.}~\bibnamefont {Babbush}},\ }\href {https://arxiv.org/abs/2312.07654}
  {\bibinfo {title} {Quantum simulation of realistic materials in first
  quantization using non-local pseudopotentials}} (\bibinfo {year} {2024}),\
  \Eprint {https://arxiv.org/abs/2312.07654} {arXiv:2312.07654 [quant-ph]}
  \BibitemShut {NoStop}%
\bibitem [{\citenamefont {Georges}\ \emph {et~al.}(2024)\citenamefont
  {Georges}, \citenamefont {Bothe}, \citenamefont {Sünderhauf}, \citenamefont
  {Berntson}, \citenamefont {Izsák},\ and\ \citenamefont
  {Ivanov}}]{georges2024}%
  \BibitemOpen
  \bibfield  {author} {\bibinfo {author} {\bibfnamefont {T.~N.}\ \bibnamefont
  {Georges}}, \bibinfo {author} {\bibfnamefont {M.}~\bibnamefont {Bothe}},
  \bibinfo {author} {\bibfnamefont {C.}~\bibnamefont {Sünderhauf}}, \bibinfo
  {author} {\bibfnamefont {B.~K.}\ \bibnamefont {Berntson}}, \bibinfo {author}
  {\bibfnamefont {R.}~\bibnamefont {Izsák}},\ and\ \bibinfo {author}
  {\bibfnamefont {A.~V.}\ \bibnamefont {Ivanov}},\ }\href
  {https://arxiv.org/abs/2408.03145} {\bibinfo {title} {Quantum simulations of
  chemistry in first quantization with any basis set}} (\bibinfo {year}
  {2024}),\ \Eprint {https://arxiv.org/abs/2408.03145} {arXiv:2408.03145
  [quant-ph]} \BibitemShut {NoStop}%
\bibitem [{\citenamefont {Berry}\ \emph {et~al.}(2018)\citenamefont {Berry},
  \citenamefont {Kieferov{\'a}}, \citenamefont {Scherer}, \citenamefont
  {Sanders}, \citenamefont {Low}, \citenamefont {Wiebe}, \citenamefont
  {Gidney},\ and\ \citenamefont {Babbush}}]{Berry2018}%
  \BibitemOpen
  \bibfield  {author} {\bibinfo {author} {\bibfnamefont {D.~W.}\ \bibnamefont
  {Berry}}, \bibinfo {author} {\bibfnamefont {M.}~\bibnamefont
  {Kieferov{\'a}}}, \bibinfo {author} {\bibfnamefont {A.}~\bibnamefont
  {Scherer}}, \bibinfo {author} {\bibfnamefont {Y.~R.}\ \bibnamefont
  {Sanders}}, \bibinfo {author} {\bibfnamefont {G.~H.}\ \bibnamefont {Low}},
  \bibinfo {author} {\bibfnamefont {N.}~\bibnamefont {Wiebe}}, \bibinfo
  {author} {\bibfnamefont {C.}~\bibnamefont {Gidney}},\ and\ \bibinfo {author}
  {\bibfnamefont {R.}~\bibnamefont {Babbush}},\ }\href
  {https://doi.org/10.1038/s41534-018-0071-5} {\bibfield  {journal} {\bibinfo
  {journal} {npj Quantum Information}\ }\textbf {\bibinfo {volume} {4}},\
  \bibinfo {pages} {22} (\bibinfo {year} {2018})}\BibitemShut {NoStop}%
\bibitem [{\citenamefont {Kitaev}(1995)}]{kitaev1995}%
  \BibitemOpen
  \bibfield  {author} {\bibinfo {author} {\bibfnamefont {A.~Y.}\ \bibnamefont
  {Kitaev}},\ }\href {https://arxiv.org/abs/quant-ph/9511026} {\bibinfo {title}
  {Quantum measurements and the abelian stabilizer problem}} (\bibinfo {year}
  {1995}),\ \Eprint {https://arxiv.org/abs/quant-ph/9511026}
  {arXiv:quant-ph/9511026 [quant-ph]} \BibitemShut {NoStop}%
\bibitem [{\citenamefont {Low}\ and\ \citenamefont {Chuang}(2019)}]{Low2019}%
  \BibitemOpen
  \bibfield  {author} {\bibinfo {author} {\bibfnamefont {G.~H.}\ \bibnamefont
  {Low}}\ and\ \bibinfo {author} {\bibfnamefont {I.~L.}\ \bibnamefont
  {Chuang}},\ }\href {https://doi.org/10.22331/q-2019-07-12-163} {\bibfield
  {journal} {\bibinfo  {journal} {{Quantum}}\ }\textbf {\bibinfo {volume}
  {3}},\ \bibinfo {pages} {163} (\bibinfo {year} {2019})}\BibitemShut {NoStop}%
\bibitem [{\citenamefont {Kivlichan}\ \emph {et~al.}(2020)\citenamefont
  {Kivlichan}, \citenamefont {Gidney}, \citenamefont {Berry}, \citenamefont
  {Wiebe}, \citenamefont {McClean}, \citenamefont {Sun}, \citenamefont {Jiang},
  \citenamefont {Rubin}, \citenamefont {Fowler}, \citenamefont {Aspuru-Guzik},
  \citenamefont {Neven},\ and\ \citenamefont {Babbush}}]{Kivlichan2020}%
  \BibitemOpen
  \bibfield  {author} {\bibinfo {author} {\bibfnamefont {I.~D.}\ \bibnamefont
  {Kivlichan}}, \bibinfo {author} {\bibfnamefont {C.}~\bibnamefont {Gidney}},
  \bibinfo {author} {\bibfnamefont {D.~W.}\ \bibnamefont {Berry}}, \bibinfo
  {author} {\bibfnamefont {N.}~\bibnamefont {Wiebe}}, \bibinfo {author}
  {\bibfnamefont {J.}~\bibnamefont {McClean}}, \bibinfo {author} {\bibfnamefont
  {W.}~\bibnamefont {Sun}}, \bibinfo {author} {\bibfnamefont {Z.}~\bibnamefont
  {Jiang}}, \bibinfo {author} {\bibfnamefont {N.}~\bibnamefont {Rubin}},
  \bibinfo {author} {\bibfnamefont {A.}~\bibnamefont {Fowler}}, \bibinfo
  {author} {\bibfnamefont {A.}~\bibnamefont {Aspuru-Guzik}}, \bibinfo {author}
  {\bibfnamefont {H.}~\bibnamefont {Neven}},\ and\ \bibinfo {author}
  {\bibfnamefont {R.}~\bibnamefont {Babbush}},\ }\href
  {https://doi.org/10.22331/q-2020-07-16-296} {\bibfield  {journal} {\bibinfo
  {journal} {{Quantum}}\ }\textbf {\bibinfo {volume} {4}},\ \bibinfo {pages}
  {296} (\bibinfo {year} {2020})}\BibitemShut {NoStop}%
\bibitem [{\citenamefont {Mikkelsen}\ and\ \citenamefont
  {Nakagawa}(2024)}]{mikkelsen2024}%
  \BibitemOpen
  \bibfield  {author} {\bibinfo {author} {\bibfnamefont {M.}~\bibnamefont
  {Mikkelsen}}\ and\ \bibinfo {author} {\bibfnamefont {Y.~O.}\ \bibnamefont
  {Nakagawa}},\ }\href {https://arxiv.org/abs/2412.13839} {\bibinfo {title}
  {Quantum-selected configuration interaction with time-evolved state}}
  (\bibinfo {year} {2024}),\ \Eprint {https://arxiv.org/abs/2412.13839}
  {arXiv:2412.13839 [quant-ph]} \BibitemShut {NoStop}%
\bibitem [{\citenamefont {Sugisaki}\ \emph {et~al.}(2024)\citenamefont
  {Sugisaki}, \citenamefont {Kanno}, \citenamefont {Itoko}, \citenamefont
  {Sakuma},\ and\ \citenamefont {Yamamoto}}]{sugisaki2024}%
  \BibitemOpen
  \bibfield  {author} {\bibinfo {author} {\bibfnamefont {K.}~\bibnamefont
  {Sugisaki}}, \bibinfo {author} {\bibfnamefont {S.}~\bibnamefont {Kanno}},
  \bibinfo {author} {\bibfnamefont {T.}~\bibnamefont {Itoko}}, \bibinfo
  {author} {\bibfnamefont {R.}~\bibnamefont {Sakuma}},\ and\ \bibinfo {author}
  {\bibfnamefont {N.}~\bibnamefont {Yamamoto}},\ }\href
  {https://arxiv.org/abs/2412.07218} {\bibinfo {title} {Hamiltonian
  simulation-based quantum-selected configuration interaction for large-scale
  electronic structure calculations with a quantum computer}} (\bibinfo {year}
  {2024}),\ \Eprint {https://arxiv.org/abs/2412.07218} {arXiv:2412.07218
  [quant-ph]} \BibitemShut {NoStop}%
\bibitem [{\citenamefont {Yu}\ \emph {et~al.}(2025)\citenamefont {Yu},
  \citenamefont {Moreno}, \citenamefont {Iosue}, \citenamefont {Bertels},
  \citenamefont {Claudino}, \citenamefont {Fuller}, \citenamefont
  {Groszkowski}, \citenamefont {Humble}, \citenamefont {Jurcevic},
  \citenamefont {Kirby}, \citenamefont {Maier}, \citenamefont {Motta},
  \citenamefont {Pokharel}, \citenamefont {Seif}, \citenamefont {Shehata},
  \citenamefont {Sung}, \citenamefont {Tran}, \citenamefont {Tripathi},
  \citenamefont {Mezzacapo},\ and\ \citenamefont {Sharma}}]{yu2025}%
  \BibitemOpen
  \bibfield  {author} {\bibinfo {author} {\bibfnamefont {J.}~\bibnamefont
  {Yu}}, \bibinfo {author} {\bibfnamefont {J.~R.}\ \bibnamefont {Moreno}},
  \bibinfo {author} {\bibfnamefont {J.~T.}\ \bibnamefont {Iosue}}, \bibinfo
  {author} {\bibfnamefont {L.}~\bibnamefont {Bertels}}, \bibinfo {author}
  {\bibfnamefont {D.}~\bibnamefont {Claudino}}, \bibinfo {author}
  {\bibfnamefont {B.}~\bibnamefont {Fuller}}, \bibinfo {author} {\bibfnamefont
  {P.}~\bibnamefont {Groszkowski}}, \bibinfo {author} {\bibfnamefont {T.~S.}\
  \bibnamefont {Humble}}, \bibinfo {author} {\bibfnamefont {P.}~\bibnamefont
  {Jurcevic}}, \bibinfo {author} {\bibfnamefont {W.}~\bibnamefont {Kirby}},
  \bibinfo {author} {\bibfnamefont {T.~A.}\ \bibnamefont {Maier}}, \bibinfo
  {author} {\bibfnamefont {M.}~\bibnamefont {Motta}}, \bibinfo {author}
  {\bibfnamefont {B.}~\bibnamefont {Pokharel}}, \bibinfo {author}
  {\bibfnamefont {A.}~\bibnamefont {Seif}}, \bibinfo {author} {\bibfnamefont
  {A.}~\bibnamefont {Shehata}}, \bibinfo {author} {\bibfnamefont {K.~J.}\
  \bibnamefont {Sung}}, \bibinfo {author} {\bibfnamefont {M.~C.}\ \bibnamefont
  {Tran}}, \bibinfo {author} {\bibfnamefont {V.}~\bibnamefont {Tripathi}},
  \bibinfo {author} {\bibfnamefont {A.}~\bibnamefont {Mezzacapo}},\ and\
  \bibinfo {author} {\bibfnamefont {K.}~\bibnamefont {Sharma}},\ }\href
  {https://arxiv.org/abs/2501.09702} {\bibinfo {title} {Quantum-centric
  algorithm for sample-based krylov diagonalization}} (\bibinfo {year}
  {2025}),\ \Eprint {https://arxiv.org/abs/2501.09702} {arXiv:2501.09702
  [quant-ph]} \BibitemShut {NoStop}%
\bibitem [{\citenamefont {Aspuru-Guzik}\ \emph {et~al.}(2005)\citenamefont
  {Aspuru-Guzik}, \citenamefont {Dutoi}, \citenamefont {Love},\ and\
  \citenamefont {Head-Gordon}}]{Guzik2005}%
  \BibitemOpen
  \bibfield  {author} {\bibinfo {author} {\bibfnamefont {A.}~\bibnamefont
  {Aspuru-Guzik}}, \bibinfo {author} {\bibfnamefont {A.~D.}\ \bibnamefont
  {Dutoi}}, \bibinfo {author} {\bibfnamefont {P.~J.}\ \bibnamefont {Love}},\
  and\ \bibinfo {author} {\bibfnamefont {M.}~\bibnamefont {Head-Gordon}},\
  }\href {https://doi.org/10.1126/science.1113479} {\bibfield  {journal}
  {\bibinfo  {journal} {Science}\ }\textbf {\bibinfo {volume} {309}},\ \bibinfo
  {pages} {1704} (\bibinfo {year} {2005})},\ \Eprint
  {https://arxiv.org/abs/https://www.science.org/doi/pdf/10.1126/science.1113479}
  {https://www.science.org/doi/pdf/10.1126/science.1113479} \BibitemShut
  {NoStop}%
\bibitem [{\citenamefont {Veis}\ and\ \citenamefont
  {Pittner}(2014)}]{Veis2014}%
  \BibitemOpen
  \bibfield  {author} {\bibinfo {author} {\bibfnamefont {L.}~\bibnamefont
  {Veis}}\ and\ \bibinfo {author} {\bibfnamefont {J.}~\bibnamefont {Pittner}},\
  }\href {https://doi.org/10.1063/1.4880755} {\bibfield  {journal} {\bibinfo
  {journal} {The Journal of Chemical Physics}\ }\textbf {\bibinfo {volume}
  {140}},\ \bibinfo {pages} {214111} (\bibinfo {year} {2014})},\ \Eprint
  {https://arxiv.org/abs/https://pubs.aip.org/aip/jcp/article-pdf/doi/10.1063/1.4880755/15480639/214111\_1\_online.pdf}
  {https://pubs.aip.org/aip/jcp/article-pdf/doi/10.1063/1.4880755/15480639/214111\_1\_online.pdf}
  \BibitemShut {NoStop}%
\bibitem [{\citenamefont {Lee}\ \emph {et~al.}(2023)\citenamefont {Lee},
  \citenamefont {Lee}, \citenamefont {Zhai}, \citenamefont {Tong},
  \citenamefont {Dalzell}, \citenamefont {Kumar}, \citenamefont {Helms},
  \citenamefont {Gray}, \citenamefont {Cui}, \citenamefont {Liu}, \citenamefont
  {Kastoryano}, \citenamefont {Babbush}, \citenamefont {Preskill},
  \citenamefont {Reichman}, \citenamefont {Campbell}, \citenamefont {Valeev},
  \citenamefont {Lin},\ and\ \citenamefont {Chan}}]{Lee2023}%
  \BibitemOpen
  \bibfield  {author} {\bibinfo {author} {\bibfnamefont {S.}~\bibnamefont
  {Lee}}, \bibinfo {author} {\bibfnamefont {J.}~\bibnamefont {Lee}}, \bibinfo
  {author} {\bibfnamefont {H.}~\bibnamefont {Zhai}}, \bibinfo {author}
  {\bibfnamefont {Y.}~\bibnamefont {Tong}}, \bibinfo {author} {\bibfnamefont
  {A.~M.}\ \bibnamefont {Dalzell}}, \bibinfo {author} {\bibfnamefont
  {A.}~\bibnamefont {Kumar}}, \bibinfo {author} {\bibfnamefont
  {P.}~\bibnamefont {Helms}}, \bibinfo {author} {\bibfnamefont
  {J.}~\bibnamefont {Gray}}, \bibinfo {author} {\bibfnamefont {Z.-H.}\
  \bibnamefont {Cui}}, \bibinfo {author} {\bibfnamefont {W.}~\bibnamefont
  {Liu}}, \bibinfo {author} {\bibfnamefont {M.}~\bibnamefont {Kastoryano}},
  \bibinfo {author} {\bibfnamefont {R.}~\bibnamefont {Babbush}}, \bibinfo
  {author} {\bibfnamefont {J.}~\bibnamefont {Preskill}}, \bibinfo {author}
  {\bibfnamefont {D.~R.}\ \bibnamefont {Reichman}}, \bibinfo {author}
  {\bibfnamefont {E.~T.}\ \bibnamefont {Campbell}}, \bibinfo {author}
  {\bibfnamefont {E.~F.}\ \bibnamefont {Valeev}}, \bibinfo {author}
  {\bibfnamefont {L.}~\bibnamefont {Lin}},\ and\ \bibinfo {author}
  {\bibfnamefont {G.~K.-L.}\ \bibnamefont {Chan}},\ }\href
  {https://doi.org/10.1038/s41467-023-37587-6} {\bibfield  {journal} {\bibinfo
  {journal} {Nature Communications}\ }\textbf {\bibinfo {volume} {14}},\
  \bibinfo {pages} {1952} (\bibinfo {year} {2023})}\BibitemShut {NoStop}%
\bibitem [{\citenamefont {Cederbaum}\ \emph {et~al.}(2006)\citenamefont
  {Cederbaum}, \citenamefont {Alon},\ and\ \citenamefont
  {Streltsov}}]{Cederbaum2006}%
  \BibitemOpen
  \bibfield  {author} {\bibinfo {author} {\bibfnamefont {L.~S.}\ \bibnamefont
  {Cederbaum}}, \bibinfo {author} {\bibfnamefont {O.~E.}\ \bibnamefont
  {Alon}},\ and\ \bibinfo {author} {\bibfnamefont {A.~I.}\ \bibnamefont
  {Streltsov}},\ }\bibfield  {journal} {\bibinfo  {journal} {Physical Review
  A}\ }\textbf {\bibinfo {volume} {73}},\ \href
  {https://doi.org/10.1103/physreva.73.043609} {10.1103/physreva.73.043609}
  (\bibinfo {year} {2006})\BibitemShut {NoStop}%
\bibitem [{\citenamefont {Tilly}\ \emph {et~al.}(2022)\citenamefont {Tilly},
  \citenamefont {Chen}, \citenamefont {Cao}, \citenamefont {Picozzi},
  \citenamefont {Setia}, \citenamefont {Li}, \citenamefont {Grant},
  \citenamefont {Wossnig}, \citenamefont {Rungger}, \citenamefont {Booth},\
  and\ \citenamefont {Tennyson}}]{Tilly2021}%
  \BibitemOpen
  \bibfield  {author} {\bibinfo {author} {\bibfnamefont {J.}~\bibnamefont
  {Tilly}}, \bibinfo {author} {\bibfnamefont {H.}~\bibnamefont {Chen}},
  \bibinfo {author} {\bibfnamefont {S.}~\bibnamefont {Cao}}, \bibinfo {author}
  {\bibfnamefont {D.}~\bibnamefont {Picozzi}}, \bibinfo {author} {\bibfnamefont
  {K.}~\bibnamefont {Setia}}, \bibinfo {author} {\bibfnamefont
  {Y.}~\bibnamefont {Li}}, \bibinfo {author} {\bibfnamefont {E.}~\bibnamefont
  {Grant}}, \bibinfo {author} {\bibfnamefont {L.}~\bibnamefont {Wossnig}},
  \bibinfo {author} {\bibfnamefont {I.}~\bibnamefont {Rungger}}, \bibinfo
  {author} {\bibfnamefont {G.~H.}\ \bibnamefont {Booth}},\ and\ \bibinfo
  {author} {\bibfnamefont {J.}~\bibnamefont {Tennyson}},\ }\href
  {https://doi.org/https://doi.org/10.1016/j.physrep.2022.08.003} {\bibfield
  {journal} {\bibinfo  {journal} {Physics Reports}\ }\textbf {\bibinfo {volume}
  {986}},\ \bibinfo {pages} {1} (\bibinfo {year} {2022})},\ \bibinfo {note}
  {the Variational Quantum Eigensolver: a review of methods and best
  practices}\BibitemShut {NoStop}%
\bibitem [{\citenamefont {Anand}\ \emph {et~al.}(2022)\citenamefont {Anand},
  \citenamefont {Schleich}, \citenamefont {Alperin-Lea}, \citenamefont
  {Jensen}, \citenamefont {Sim}, \citenamefont {Díaz-Tinoco}, \citenamefont
  {Kottmann}, \citenamefont {Degroote}, \citenamefont {Izmaylov},\ and\
  \citenamefont {Aspuru-Guzik}}]{Anand2022}%
  \BibitemOpen
  \bibfield  {author} {\bibinfo {author} {\bibfnamefont {A.}~\bibnamefont
  {Anand}}, \bibinfo {author} {\bibfnamefont {P.}~\bibnamefont {Schleich}},
  \bibinfo {author} {\bibfnamefont {S.}~\bibnamefont {Alperin-Lea}}, \bibinfo
  {author} {\bibfnamefont {P.~W.~K.}\ \bibnamefont {Jensen}}, \bibinfo {author}
  {\bibfnamefont {S.}~\bibnamefont {Sim}}, \bibinfo {author} {\bibfnamefont
  {M.}~\bibnamefont {Díaz-Tinoco}}, \bibinfo {author} {\bibfnamefont {J.~S.}\
  \bibnamefont {Kottmann}}, \bibinfo {author} {\bibfnamefont {M.}~\bibnamefont
  {Degroote}}, \bibinfo {author} {\bibfnamefont {A.~F.}\ \bibnamefont
  {Izmaylov}},\ and\ \bibinfo {author} {\bibfnamefont {A.}~\bibnamefont
  {Aspuru-Guzik}},\ }\href {https://doi.org/10.1039/D1CS00932J} {\bibfield
  {journal} {\bibinfo  {journal} {Chem. Soc. Rev.}\ }\textbf {\bibinfo {volume}
  {51}},\ \bibinfo {pages} {1659} (\bibinfo {year} {2022})}\BibitemShut
  {NoStop}%
\bibitem [{\citenamefont {Horiba}\ \emph {et~al.}(2023)\citenamefont {Horiba},
  \citenamefont {Shirai},\ and\ \citenamefont {Hirai}}]{horiba2023}%
  \BibitemOpen
  \bibfield  {author} {\bibinfo {author} {\bibfnamefont {T.}~\bibnamefont
  {Horiba}}, \bibinfo {author} {\bibfnamefont {S.}~\bibnamefont {Shirai}},\
  and\ \bibinfo {author} {\bibfnamefont {H.}~\bibnamefont {Hirai}},\ }\href
  {https://arxiv.org/abs/2306.08434} {\bibinfo {title} {Construction of
  antisymmetric variational quantum states with real-space representation}}
  (\bibinfo {year} {2023}),\ \Eprint {https://arxiv.org/abs/2306.08434}
  {arXiv:2306.08434 [quant-ph]} \BibitemShut {NoStop}%
\bibitem [{\citenamefont {Mikkelsen}\ and\ \citenamefont
  {Danshita}(2025)}]{Mikkelsen2025}%
  \BibitemOpen
  \bibfield  {author} {\bibinfo {author} {\bibfnamefont {M.}~\bibnamefont
  {Mikkelsen}}\ and\ \bibinfo {author} {\bibfnamefont {I.}~\bibnamefont
  {Danshita}},\ }\href {https://arxiv.org/abs/2510.25358} {\bibinfo {title}
  {Entanglement-enhanced correlation propagation in the one-dimensional su($n$)
  fermi-hubbard model}} (\bibinfo {year} {2025}),\ \Eprint
  {https://arxiv.org/abs/2510.25358} {arXiv:2510.25358 [cond-mat.quant-gas]}
  \BibitemShut {NoStop}%
\bibitem [{\citenamefont {Berry}\ \emph {et~al.}(2019)\citenamefont {Berry},
  \citenamefont {Gidney}, \citenamefont {Motta}, \citenamefont {McClean},\ and\
  \citenamefont {Babbush}}]{Berry2019}%
  \BibitemOpen
  \bibfield  {author} {\bibinfo {author} {\bibfnamefont {D.~W.}\ \bibnamefont
  {Berry}}, \bibinfo {author} {\bibfnamefont {C.}~\bibnamefont {Gidney}},
  \bibinfo {author} {\bibfnamefont {M.}~\bibnamefont {Motta}}, \bibinfo
  {author} {\bibfnamefont {J.~R.}\ \bibnamefont {McClean}},\ and\ \bibinfo
  {author} {\bibfnamefont {R.}~\bibnamefont {Babbush}},\ }\href
  {https://doi.org/10.22331/q-2019-12-02-208} {\bibfield  {journal} {\bibinfo
  {journal} {{Quantum}}\ }\textbf {\bibinfo {volume} {3}},\ \bibinfo {pages}
  {208} (\bibinfo {year} {2019})}\BibitemShut {NoStop}%
\bibitem [{qur(2022)}]{quri-parts}%
  \BibitemOpen
  \href {https://github.com/QunaSys/quri-parts} {\bibinfo {title} {{QURI
  Parts}}} (\bibinfo {year} {2022}),\ \bibinfo {note}
  {\url{https://github.com/QunaSys/quri-parts}}\BibitemShut {NoStop}%
\bibitem [{\citenamefont {Akahoshi}\ \emph {et~al.}(2024)\citenamefont
  {Akahoshi}, \citenamefont {Maruyama}, \citenamefont {Oshima}, \citenamefont
  {Sato},\ and\ \citenamefont {Fujii}}]{Akahoshi2024}%
  \BibitemOpen
  \bibfield  {author} {\bibinfo {author} {\bibfnamefont {Y.}~\bibnamefont
  {Akahoshi}}, \bibinfo {author} {\bibfnamefont {K.}~\bibnamefont {Maruyama}},
  \bibinfo {author} {\bibfnamefont {H.}~\bibnamefont {Oshima}}, \bibinfo
  {author} {\bibfnamefont {S.}~\bibnamefont {Sato}},\ and\ \bibinfo {author}
  {\bibfnamefont {K.}~\bibnamefont {Fujii}},\ }\href
  {https://doi.org/10.1103/PRXQuantum.5.010337} {\bibfield  {journal} {\bibinfo
   {journal} {PRX Quantum}\ }\textbf {\bibinfo {volume} {5}},\ \bibinfo {pages}
  {010337} (\bibinfo {year} {2024})}\BibitemShut {NoStop}%
\bibitem [{\citenamefont {{Qiskit contributors}}(2023)}]{Qiskit}%
  \BibitemOpen
  \bibfield  {author} {\bibinfo {author} {\bibnamefont {{Qiskit
  contributors}}},\ }\href {https://doi.org/10.5281/zenodo.2573505} {\bibinfo
  {title} {Qiskit: An open-source framework for quantum computing}} (\bibinfo
  {year} {2023})\BibitemShut {NoStop}%
\bibitem [{\citenamefont {Childs}\ \emph {et~al.}(2021)\citenamefont {Childs},
  \citenamefont {Su}, \citenamefont {Tran}, \citenamefont {Wiebe},\ and\
  \citenamefont {Zhu}}]{Childs2021}%
  \BibitemOpen
  \bibfield  {author} {\bibinfo {author} {\bibfnamefont {A.~M.}\ \bibnamefont
  {Childs}}, \bibinfo {author} {\bibfnamefont {Y.}~\bibnamefont {Su}}, \bibinfo
  {author} {\bibfnamefont {M.~C.}\ \bibnamefont {Tran}}, \bibinfo {author}
  {\bibfnamefont {N.}~\bibnamefont {Wiebe}},\ and\ \bibinfo {author}
  {\bibfnamefont {S.}~\bibnamefont {Zhu}},\ }\href
  {https://doi.org/10.1103/PhysRevX.11.011020} {\bibfield  {journal} {\bibinfo
  {journal} {Phys. Rev. X}\ }\textbf {\bibinfo {volume} {11}},\ \bibinfo
  {pages} {011020} (\bibinfo {year} {2021})}\BibitemShut {NoStop}%
\end{thebibliography}%

\onecolumngrid
\renewcommand{\thefigure}{S\arabic{figure}}
\renewcommand{\thetable}{S\arabic{table}}
\renewcommand{\theequation}{S\arabic{equation}}
\setcounter{equation}{0}
\setcounter{figure}{0}
\setcounter{table}{0}
\FloatBarrier
\begin{appendix}
\newpage
\begin{center}
{
\large
Supplemental Material: ``First and second quantized digital quantum simulations of bosonic systems''
}
\end{center}

\section{Comparing with optimized CNOT gate counts}
\label{sec:appendixA}
\begin{figure*}[htb!]
\centering
\includegraphics[width=1.0\linewidth]{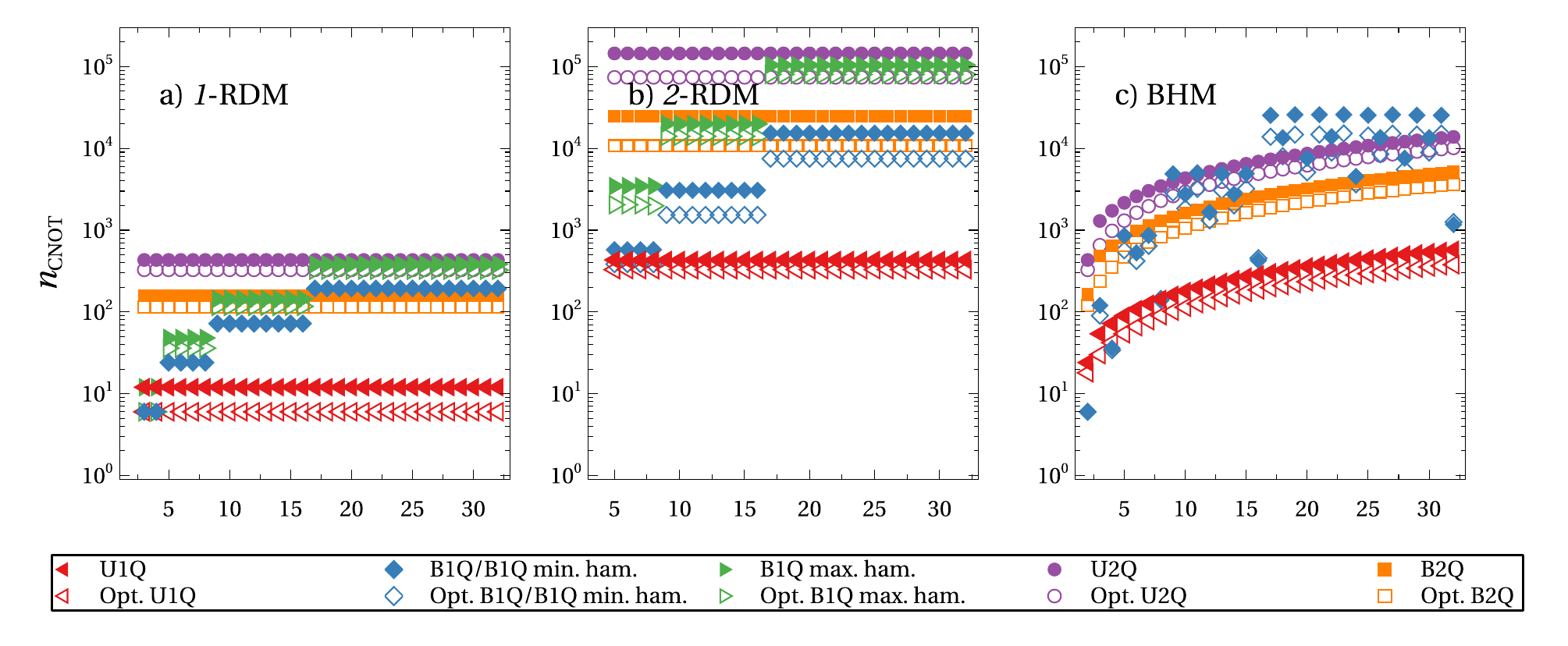}
\caption{Comparing the number of CNOT gates required to express the exponential for (a) the $1$-RDM ODTs, (b) the $2$-RDM ODTs and (c) the BHM as a function of $M$. In addition to the raw CNOT gate counts the same figure after running the Qiskit optimizer is displayed by the unfilled symbols.}
\label{fig:Qiskitoptimizationcomparison}  
\end{figure*} 

In the main text we considered CNOT gates based on the basic CNOT staircase approach without further simplifications applied to the circuit. The optimization routines of the Qiskit software \cite{Qiskit}, can reduce the number of CNOT gates by e.g. replacing two subsequent CNOT gates with an identity etc. We use the Qiskit optimizer with the optimization level set to 3. For the $1$-RDM, a minimum Hamming distance of 1, corresponding to nearest-neighbor hopping, does not benefit from Qiskit optimization, as it is already optimal. Similarly the BHM with $M=2^n$, does not benefit from the optimization, suggesting that these values are already quite optimized. Aside from these special cases, the Qiskit optimization can generally improve the CNOT gate counts by a factor of 2-3, but without changing the relative efficiency of the first and second quantized mappings.  
\section{Analyzing Trotter errors at the operator level}
\label{sec:appendixB}
In the main text we investigated the resource cost per trotter step for the BHM and HO, but When implementing the time-evolution operator the number of steps required to obtain a desired accuracy is also important. While each mapping represents the same Hamiltonian, the different qubit representations of the Hamiltonian can potentially lead to different Trotter errors. In order to evaluate the trotter error in principle one has to evaluate an infinite sum of nested commutators, so in practice some approximation for bounding the error is usually done. For a Hamiltonian given as a linear sum of Paulis strings $H=\sum_{l}w_l P_l$ the first-order Trotter approximation using $n_\text{steps}$ is given by
\begin{align}
S_1(t)=\left(\prod_l e^{-i w_l \hat{P}_l \Delta t}\right)^{n_{\text{step}}}+O((\Delta t)^2).
\end{align}
In the literature a naive estimate of the second-order Trotter error is bound by \cite{Childs2021}
\begin{align}
||S_1(t) - e^{-itH}|| \le \frac{t^2}{2} \sum_{\gamma_1=1}^\Gamma \sum_{\gamma_2=\gamma_1+1}^\Gamma \left\| [ P_{\gamma_2}, P_{\gamma_1} ] \right\|
\end{align}
that is the sum over the spectral norm of the pairwise sum of the commutators of each string. The evaluation of this scales roughly as $\mathcal{O}(\Gamma^2 n)$, where $\Gamma$ is the number of pauli strings and $n$ is the number of qubits. This estimate is known to strongly overestimate the trotter error in many cases and a more sophisticated bound for the second-order Trotter error can be obtained using a grouped commutator \cite{Childs2021}
\begin{align}
||S_1(t) - e^{-itH}|| \le \frac{t^2}{2} \sum_{\gamma_1=1}^\Gamma \left\| \left[ \sum_{\gamma_2=\gamma_1+1}^\Gamma P_{\gamma_2}, P_{\gamma_1} \right] \right\|    
\end{align}
where the norm is the spectral norm of the grouped commutator. This allows for cancellations to occur and therefore it provides a tighter bound, although it still overestimates compared to the actual error. However, this scales as $\mathcal{O}(\Gamma^2 n + \Gamma \cdot 2^n)$, because a SVD is required to calculate the spectral norm of the grouped commutator. If using optimized sparse SVD routines, the $\Gamma^2 n$ cost can be worse for small systems, but it is the exponential cost $\Gamma \cdot 2^n$ that makes this error bound intractable for larger systems, while the simpler bound can in principle be calculated.

Finally we can calculate the spectral norm 
\begin{align}
||S_1(t) - e^{-itH}||     
\end{align}
directly. This requires evaluating the full matrix exponential which scales as roughly $2^{3n}$ for dense exponentiation which is obviously intractable for large systems. While the computational complexity can be somewhat improved, in particular by considering expectation values of representative states which allows one to take advantage of sparsity this is beyond the scope of this section. 

As we have shown that the number of Pauli strings scale worse with system size for the second quantized mappings, the most naive error bound clearly scales worse as it is proportional to the pair of all commutators. Indeed investigating the BHM for small $N,M$ as seen in table \ref{tab:trotter_errors} this is borne out. The more sophisticated error bound based on the grouped commutator is more similar across the mappings, but still somewhat larger for the second quantized ones. Interestingly the actual error is quite similar for all the mappings. This suggests that any attempt to derive trends based on the naive bound for larger systems is likely fallacious. Overall the actual errors suggest that the per-step Trotter cost is the important signifier of the difference between the mappings, but we note that having a tighter bound based on the approximate calculations is still useful in order to know how tight of an approximation to implement in practice. The difficulty of tight error estimates for large systems is one of the shortcomings of trotterization. Using quibitization the number of circuit queries has a linear dependence on the simple one-norm of the Hamiltonian which can easily be calculated regardless of system-size which is one of the advantages of the latter method. Therefore the 1-norm/qubitization comparison in the main text is indeed a more robust comparison of the mappings for specific Hamiltonians.

\begin{table*}[htb!]
\centering
\renewcommand{\arraystretch}{1.2}
\begin{tabular}{l|c|c|c|c|c|c|c|}
\hline
\textbf{System size} & \textbf{N=3, M=3} & \textbf{N=3, M=4} & \textbf{N=3, M=5} & \textbf{N=4, M=3} & \textbf{N=4, M=4} & \textbf{N=4, M=5} & HO \textbf{N=3, M=4} \\
\hline
\multicolumn{8}{c}{\textbf{U1Q}} \\
\hline
Pairwise error bound & $4.05$ & $5.40$ & $6.75$ & $7.80$ & $11.4$ & $13.0 $ & $15.5$ \\
Grouped error bound & $1.85$ & $2.47$ & $3.09$ & $3.65$ & $4.87$ & $6.08$ & $3.98$\\
Actual error & $2.33 \cdot 10^{-1}$ & $3.49 \cdot 10^{-1}$ & $2.78 \cdot 10^{-1}$ & $3.08 \cdot 10^{-1}$ & $4.79 \cdot 10^{-1}$ & $3.84 \cdot 10^{-1}$ & $1.80 \cdot 10^{-1}$ \\
\hline
\multicolumn{8}{c}{\textbf{B1Q}} \\
\hline
Pairwise error bound & $3.15$ & $1.20$ & $5.10$ & $6.00$ & $2.40$ & $9.60$ & $2.48$ \\
Grouped error bound & $1.85$ & $1.20$ & $2.50$ & $3.65$ & $2.40$ & $4.89$ & $1.20$\\
Actual error & $2.33 \cdot 10^{-1}$ & $2.43 \cdot 10^{-1}$ & $2.90 \cdot 10^{-1}$ & $3.08 \cdot 10^{-1}$ & $3.89 \cdot 10^{-1}$ & $3.90 \cdot 10^{-1}$ & $1.33 \cdot 10^{-1}$ \\
\hline
\multicolumn{8}{c}{\textbf{U2Q}} \\
\hline
Pairwise error bound & $74.7$ & $99.6 $ & $125$ & $323$ & $431$ & $539$ & $4327$ \\
Grouped error bound & $4.70$ & $6.16$ & $7.74$ & $12.0 $ & $15.7 $ & $19.7$ & $20.5$ \\
Actual error & $3.23 \cdot 10^{-1}$ & $3.71 \cdot 10^{-1}$ & $3.55 \cdot 10^{-1}$ & $4.67 \cdot 10^{-1}$ & $5.11 \cdot 10^{-1}$ & $4.90 \cdot 10^{-1}$ & $5.27 \cdot 10^{-1}$ \\
\hline
\multicolumn{8}{c}{\textbf{B2Q}} \\
\hline
Pairwise error bound & $34.7$ & $46.3$ & $57.9$ & $229$ & $305$ & $381$ &$246$ \\
Grouped error bound & $5.67$ & $7.40$ & $9.33$ & $13.7$ & $17.8 $ & $22.4$ & $9.49$ \\
Actual error & $2.71 \cdot 10^{-1}$ & $3.33 \cdot 10^{-1}$ & $3.18 \cdot 10^{-1}$ & $3.80 \cdot 10^{-1}$ & $4.73 \cdot 10^{-1}$ & $4.36 \cdot 10^{-1}$ & $2.30 \cdot 10^{-1}$ \\
\hline
\end{tabular}
\caption{First-order Trotter error bounds and actual errors for $t=1.0$, and  $r=10$ steps. The last column corresponds to the harmonic oscillator with $g=2$, while all other columns correspond to the BHM with $U=2J$.}
\label{tab:trotter_errors}
\end{table*}
\end{appendix}

\end{document}